\documentclass[preprint,english,3p,times,sort&compress,procedia]{elsarticle}
\usepackage{amssymb} 
\usepackage{amsthm} 

\usepackage[ansinew]{inputenc} 
\usepackage[english]{babel}    
\usepackage{amsmath}           
\usepackage{amsfonts}
\usepackage{latexsym}
\usepackage{multirow}           
\usepackage{array}              
\usepackage{longtable}          
\usepackage{graphicx}           
\usepackage{subfigure}          
\usepackage{pstricks,pst-node}           
\usepackage{natbib}             
\usepackage{wrapfig}            
\usepackage{hyperref}           

\begin{document}

\begin{frontmatter}




\title{Two-level Continuous Topology Optimization in Structural Mechanics}

\author[1,2]{Rafael Merli\corref{cor1}}
\cortext[cor1]{Corresponding author at: Universitat Politècnica de València, Departamento de Ingeniería Mecánica y de Materiales. Camino de vera s/n, 46022 Valencia, Spain. Tel:+34 96 387 76 21. Fax: +34 96 387 76 29}
\ead{ramergis@upvnet.upv.es}

\author[1]{Antolín Martínez-Martínez}

\author[1]{Juan José Ródenas}

\author[1]{Marc Bosch-Galera}

\author[1]{Enrique Nadal}

\address[1]{Departamento de Ingeniería Mecánica y de Materiales.\\ Escuela Técnica Superior de Ingenieros Industriales.\\ Universitat Politècnica de València. Camino de vera s/n, 46022 Valencia, Spain}
\address[2]{Centro Universitario EDEM - Escuela de Empresarios.\\ Muelle de la Aduana s/n. La Marina de Valencia, 46024 Valencia, Spain}

\begin{abstract}
In the current industry, the development of optimized mechanical components able to satisfy the customer requirements evolves quickly. Therefore, companies are asked for efficient solutions to improve their products in terms of stiffness and strength. In this sense, Topology Optimization has been extensively used to determine the best topology of structural components from the mechanical point of view. Its main objective is to distribute a given amount of material into a predefined domain to reach the maximum overall stiffness of the component. Besides, high-resolution solutions are essential to define the final distribution of material. Standard Topological Optimization tools are able to propose an optimal topology for the whole component, but when small topological details are required (i.e. trabecular-type structures) the computational cost is prohibitive. In order to mitigate this issue, the present work proposes a two-level topology optimization method to solve high-resolution problems by using density-based methods. The proposed methodology includes three steps: The first one subdivides the whole component in cells and generates a coarse optimized low-definition material distribution assigning one different density to each cell. The second one uses an equilibrating technique that provides tractions continuity between adjacent cells, thus ensuring the material inter-cell continuity after the cells optimization process. Finally, each cell is optimized at fine scale taking as input data the densities and the equilibrated tractions obtained from the macro problem. The main goal of this work is to efficiently solve high-resolution topology optimization problems using density-based methods, which would be unaffordable with standard computing facilities and the current methodologies.
\end{abstract}


\end{frontmatter}


\section{Introduction}\label{Introd}
The optimization of mechanical components represents a major challenge in scientific and engineering fields. Before the huge development of computer science, structural optimization was the result of a long trial and error process. In fact, skills and intuition of engineers strongly conditioned the final designs. However, the increasing economical, industrial and social requirements need automated processes to develop optimized structures more rapidly than by traditional procedures.\par

In fact, numerical simulation of structural problems can considerably help to take better decisions along the design phase of complex structures. These numerical methods, e.g. the Finite Element Method (FEM), allow us to know an accurate prediction of the structural response of a component under the prescribed external loads. In fact, the FEM has streamlined remarkably the design phase.\par

Despite the huge advance of numerical methods, the process to distribute the material into a design domain is not completely automated, but is highly dependend on the decisions made by the engineer. The following step in this regard is the application of systematic techniques for structural optimization, taking advantage of the FEM, not only to evaluate the structural behavior, but also to propose new configurations in the material distribution.\par

In the structural optimization field, three main groups not necessarily separated of optimization problems \cite{Eschenauer2001} can be found. Leaving aside the size optimization, which is not directly related to this work, the most relevant are shape optimization and topology optimization. For instance, in \cite{Marco2018} a base structure is defined in terms of parameters related with dimensions (like diameter, section's height, etc). Thus, the algorithm will decide the value of each parameter of the structure in order to get the higher stiffness with the minimum required material, while preserving the connectivity of the structure. An alternative shape optimization technique is the so-called \emph{bubble method} \cite{Eschenauer1994}, where new holes (``bubbles") are inserted or removed at determinate points in the domain and shape optimization is carried out simultaneously.\par

The present paper is focused on the third group of optimization problems, where the topological optimization objective consists on obtaining an appropriate distribution of certain amount of material into a predefined design domain, to minimize the compliance of the component under a set of predefined loads. As usual, it involves defining a fixed finite element (FE) mesh along the whole domain. In fact, the quality of the final resolution is highly conditioned by the element size of the mesh.\par

Many structures optimized by nature present trabecular topologies (e.g. bone tissues) which properly work with lightweight material distributions. In fact, it has been proved that these trabecular structures are versatile to withstand load variations \cite{Wu2021}, different from those set in the initial design. Thus, it would be highly desirable to obtain trabecular structures, described with high resolution, as the solution of topology optimization problems. Furthermore, they are also more tolerant to bar failures since local buckling can be controlled directly by the cell size and the tissue thickness inside each cell. Namely, it is a buckling control by construction.\par

The first approach is using an extremely refined FE meshes and adapt classical topological optimization techniques, as the well-known Solid Isotropic Material with Penalization (SIMP) \cite{Bendsoe2003, Bendsoe1999}, to solve very large domains. However, this would involve a high computational cost. On the other hand, the multi-scale strategies use local results from a coarse scale FEM model to determine the solution into each element at the fine scale.\par

Included in multi-scale methods, \cite{ZWu2019} uses predefined lattice structures (square frame and frame with crossed trusses) into each element and relates the density of the cell with the feature's thickness. Hence, the final geometry is determined by the coarse scale minimization of structural compliance subject to a volume fraction constraint. Although the lattice structures can be defined to reach a continuous solution, the final solution will be far from being fully optimized because the topology of the cells is predefined and, therefore, not optimized.\par

Alternatively to these lattice structures, the work presented in \cite{Ferrer2016} deals with two different levels (coarse and fine structure) in the resolution process, which involves a homogenization step in the fine scale to define an equivalent material at each integration point and solve again the coarse problem up to convergence. Since this procedure is computationally expensive, the authors propose a new strategy based on generating a database of fine-scale solutions, each of them corresponding to a single macroscopic stress state. Joining afterwards these cells, the entire design domain could be optimized approximately. However, the solution obtained by this method is not structurally viable because continuity between cells has not been taken into account.\par

The main goal of this work is to develop a two-level topology optimization method which provides high-resolution images with fully optimized cells and mechanical continuity between adjacent cells. This continuity makes the resulting component an ideal candidate to be additive manufactured \cite{Panesar2018}. Although herein it is assumed that each manufacturable cell only involves one coarse element for simplicity, each of the cells could also be defined by a set of neighoburing coarse elements.\par

The paper is organized as follows: Section \ref{MainMet} summarizes the basics of 2D optimization problems employed herein. In Section \ref{2level} the proposed two-level optimization methodology and its numerical implementation are presented. Sample problems and their numerical results compared with some published references are presented in Section \ref{results}. Finally, concluding remarks are addressed in Section \ref{remarks}.

\section{Main Methodologies}\label{MainMet}
\subsection{Two-dimensional Linear Elasticity Problem}\label{Lproblem}
The general 2D linear elasticity problem and the notation used hereafter are described next.\par
 \begin{figure}[h! t]
\begin{center}
    \includegraphics[scale=0.7]{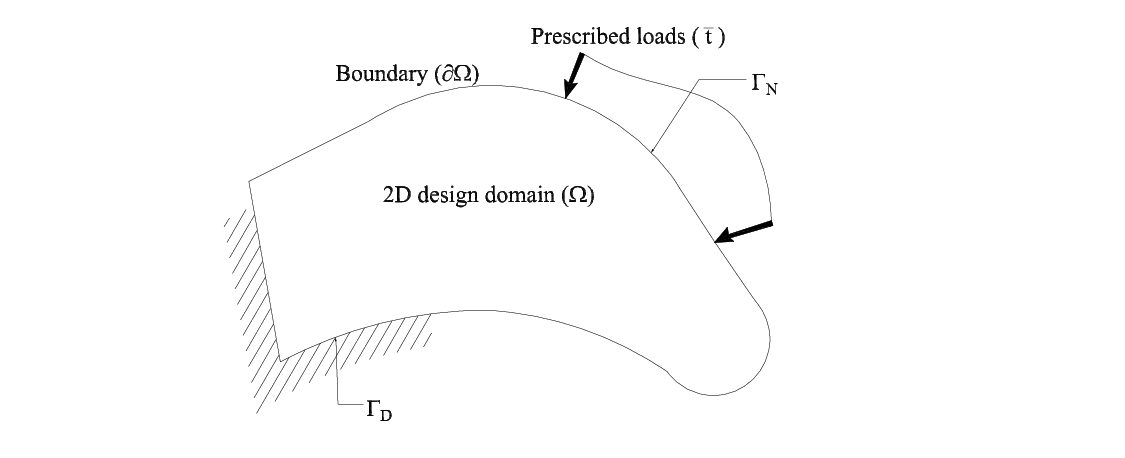}
\end{center}
\caption{2D design problem}\label{fig1}
\end{figure}

We seek a displacement field \textbf{u} and a Cauchy stress field $\boldsymbol{\sigma}$ defined over the domain $\Omega$ (Figure \ref{fig1}) verifying:
\begin{itemize}
  \item Static admissibility, involving internal equilibrium and Neumann boundary conditions (BC)
    \begin{subequations}\label{eq1}
      \begin{align}
           & \mathbf{L}^T \boldsymbol{\sigma}+\mathbf{b} = \mathbf{0} & & \text{in}\ \Omega \label{eq1a}\\
           & \mathbf{G} \boldsymbol{\sigma} = \mathbf{\bar{t}} & & \text{in}\ \Gamma_N,  \label{eq1b}
      \end{align}
    \end{subequations}
    where $\mathbf{L}$ is the differential operator defined by:
    \begin{equation}\label{eq2}
        \mathbf{L}= \begin{bmatrix}
                     \frac{\partial}{\partial x} & 0 \\
                     0 & \frac{\partial}{\partial y} \\
                     \frac{\partial}{\partial y} & \frac{\partial}{\partial x} \\
                   \end{bmatrix}
    \end{equation}
    and $\mathbf{G}$ is the operator which projects the stress field $\boldsymbol{\sigma}=\{\sigma_x,\sigma_y,\tau_{xy}\}^T$ into tractions over the boundary. Considering the external unit vector $\mathbf{n}=\{n_x,n_y\}^T$ normal to the boundary $\partial\Omega$ with Neumann BC $\Gamma_N$, then:
    \begin{equation}\label{eq3}
        \mathbf{G}=\begin{bmatrix}
                           n_x & 0 & n_y\\
                           0 & n_y & n_x\\
                         \end{bmatrix}.
    \end{equation}
  \item Dirichlet BC on $\Gamma_D \subset \partial\Omega$\ (note that $\Gamma_D \cap \Gamma_N=\emptyset$)
      \begin{equation}\label{eq4}
        \mathbf{u}=\mathbf{\bar{u}}.
      \end{equation}
  \item Constitutive relations:
      \begin{equation}\label{eq5}
        \boldsymbol{\sigma}=\mathbf{D}\boldsymbol{\varepsilon}(\mathbf{u}),
      \end{equation}
      where $\boldsymbol{\varepsilon}(\mathbf{u})=\mathbf{L}\mathbf{u}=\{\varepsilon_x,\varepsilon_y,\gamma_{xy}\}^T$ and $\mathbf{D}$ contains the mechanical parameters of the elastic isotropic material.
\end{itemize}
Alternatively, the problem above can be expressed in variational form by defining:
\begin{align}\label{eq7}
    \alpha(\mathbf{u},\mathbf{v})=\int_\Omega \boldsymbol\varepsilon(\mathbf{u})^T\mathbf{D}\boldsymbol\varepsilon(\mathbf{v}) d\Omega &  & l(\mathbf{v})=\int_\Omega\mathbf{b}^T\mathbf{v}d\Omega+\int_{\Gamma_N}\mathbf{\bar{t}}^T\mathbf{v}d\Gamma_N,
\end{align}
thereby, we should:
\begin{align}
     & \text{Find}\  \{\mathbf{u} \in (V\cup\{\mathbf{w}\}) / \alpha(\mathbf{u},\mathbf{v})=l(\mathbf{v})\} \hspace{20pt}  \forall v \in V,\label{eq8}\\
     & \text{where}\  V=\{\mathbf{v}\in \mathcal{C}^1(\Omega)/\mathbf{v}|_{\Gamma_D}=\mathbf{0}\} \label{eq9}
\end{align}
and $\mathbf{w}$ is a particular displacement field satisfying the Dirichlet BC \eqref{eq4}.

\subsection{Topology Optimization Problem}\label{TOproblem} 
Following \cite{Bendsoe2003}, the SIMP method can be adopted to solve the topology optimization problem for minimum compliance, which takes the form:
\begin{equation}\label{eq10}
  \text{Find} \biggl\{ \min_{\rho \in [\rho_{min},1]} \alpha(\rho; \mathbf{u},\mathbf{u}) \biggr\},
\end{equation}
subjected to:
\begin{subequations}\label{eq11}
    \begin{gather}
         l(\mathbf{u})=\alpha(\rho; \mathbf{u},\mathbf{u}) \label{eq11a}\\
         \int_V \rho(\mathbf{x})dV = V_0, \ \forall \mathbf{x} \in \Omega, \label{eq11b}
    \end{gather}
\end{subequations}
where:\par
\begin{tabular}{l l}
     $\mathbf{u}$ & unknown displacements\\
     $V_0$ & initial volume value to be distributed in the domain $\Omega$\\
     $\rho (\mathbf{x})$ & relative density function of the position $\mathbf{x}$\\
     $\rho_{min}=10^{-3}$ & low threshold of density introduced herein to avoid any singularity in the equilibrium problem.\\
\end{tabular}\\

The constitutive matrix $\mathbf{D}$ is redefined with respect to that of a reference material $\mathbf{D}_0$, as follows:
\begin{equation}\label{eq12}
    \mathbf{D}(\mathbf{x})= \rho^p(\mathbf{x}) \mathbf{D}_0,
\end{equation}
where \(p \geq 1\) is a penalty parameter which penalizes intermediate densities (usually \(p=3\)). In this way, the functional $\alpha(\mathbf{u},\mathbf{v})$ in \eqref{eq7} becomes
\begin{equation}\label{eq13}
    \alpha(\rho; \mathbf{u},\mathbf{u})=\int_\Omega \rho^p \boldsymbol\varepsilon(\mathbf{u})^T\mathbf{D}_0 \boldsymbol\varepsilon(\mathbf{u}) d\Omega.
\end{equation}\par

\subsubsection{Optimality Criteria}\label{OCriteria}
Next, the optimality criteria for density required in the topology optimization problem defined in Section \ref{TOproblem} is outlined (see \cite{Bendsoe2003}). By applying Lagrange multipliers to constraints \eqref{eq11}, the constraint for $\rho$ becomes:
\begin{equation}\label{eq16}
    \boldsymbol{\varepsilon}^T\frac{\partial \mathbf{D}}{\partial \rho}\boldsymbol{\varepsilon}=\Lambda+\lambda^+(\mathbf{x})-\lambda^-(\mathbf{x}),
\end{equation}
where $\Lambda$ is the multiplier for constraint \eqref{eq11b} and $\lambda^+(\mathbf{x})$ and $\lambda^-(\mathbf{x})$ are the respective multipliers for the higher and lower density limits involved in \eqref{eq11a} . For intermediate densities and using \eqref{eq12}, we obtain
\begin{equation}\label{eq17}
    p\rho(\mathbf{x})^{p-1} \boldsymbol{\varepsilon}^T \mathbf{D}_0\boldsymbol{\varepsilon}=\Lambda,
\end{equation}
which indicates that the strain energy density is constant for all intermediate densities. Thus, in order to make possible the density updating, a new design variable is defined in iteration $k$ as
\begin{equation}\label{eq18}
    B_k=\frac{1}{\Lambda_k} p\rho(\mathbf{x})^{p-1} \boldsymbol{\varepsilon}^T(\mathbf{u}_k) \mathbf{D}_0\boldsymbol{\varepsilon}(\mathbf{u}_k).
\end{equation}

Hence, the following density updating for each element is adopted \cite{Bendsoe2003}:
\begin{equation}\label{eq20}
    \rho_{k+1}^e=\begin{cases}
                \max\{(1-\zeta)\rho_k^e,\rho_{min}\} & \mbox{if}\ \rho_k^e B_k^\eta \leq \max\{(1-\zeta)\rho_k^e,\rho_{min}\}\\
                \rho_k^e B_k^\eta &  \mbox{if}\ \max\{(1-\zeta)\rho_k^e,\rho_{min}\} \leq \rho_k B_k^\eta \leq \min\{(1+\zeta)\rho_k^e,1\}\\
                \min \{(1+\zeta)\rho_k^e,1\} & \mbox{if}\ \rho_k^e B_k^\eta \geq \min\{(1+\zeta)\rho_k^e,1\}
                                      \end{cases},
\end{equation}
being $\zeta=0.2$ and $\eta=0.5$ the standard values of the parameters used to control the process. From \eqref{eq18}, a local optimum is reached if $B_k=1$ for densities $\rho_{min}<\rho<1$. In addition, the updating scheme \eqref{eq20} adds material to areas where $B_k>1$, related to high variations of strain energy density in the numerator of \eqref{eq18}. Otherwise, material is removed from areas where $B_k<1$, namely, with low variations of strain energy density.

\section{The Proposed Two-level Optimization Technique}\label{2level}
\subsection{General Procedure}\label{GeneralP}
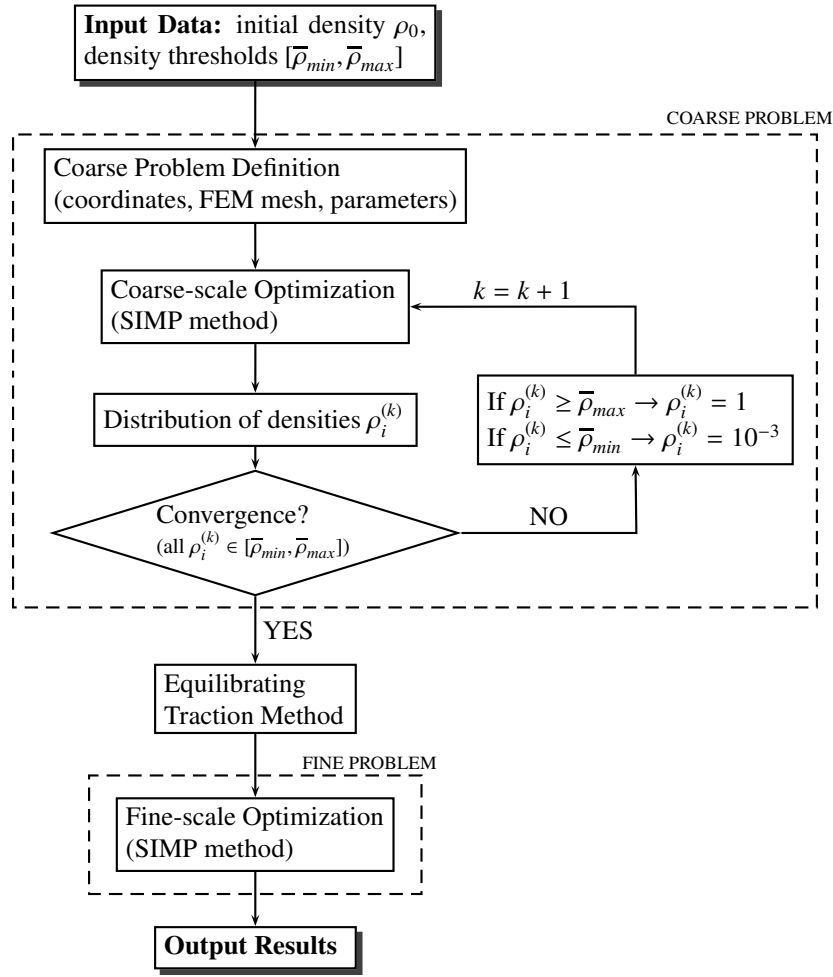
\begin{figure}[h! t]
\begin{center}
 \begin{pspicture}(0,0)(12,13)
   \rput(4,12.5){\rnode{InData}{\psframebox[shadow=true]{\parbox[c]{4.5cm}{\textbf{Input Data:} initial density $\rho_0$, density thresholds \( [\overline \rho_{min},\overline \rho_{max}]\)}}}}
   \rput(4,10.6){\rnode{MacroDef}{\psframebox{\parbox[c]{5.3cm}{Coarse Problem Definition\\ (coordinates, FEM mesh, parameters)}}}}
   \rput(4,9){\rnode{MacroOpt}{\psframebox{\parbox[c]{3.8cm}{Coarse-scale Optimization\\ (SIMP method)}}}}
   \rput(4,7.5){\rnode{DistribDens}{\psframebox{\parbox[c]{4cm}{Distribution of densities $\rho_i^{(k)}$}}}}
   \rput(4,6){\rnode{Conv}{\psdiabox[framesep=0.05]{\parbox[c]{2.6cm}{Convergence?\\ \footnotesize{(all $\rho_i^{(k)} \in [\overline \rho_{min},\overline \rho_{max}]$)}}}}}
   \rput(9,7.5){\rnode{UpdateDens}{\psframebox{\parbox[c]{3.9cm}{If $\rho_i^{(k)} \geq \overline\rho_{max} \rightarrow{\rho_i^{(k)}=1}$\\ If $\rho_i^{(k)} \leq \overline\rho_{min} \rightarrow \rho_i^{(k)}=10^{-3}$}}}}
   \rput(4,3.8){\rnode{EqTrac}{\psframebox{\parbox[c]{2.4cm}{Equilibrating Traction Method}}}}
   \rput(4,2){\rnode{MicroOpt}{\psframebox{\parbox[c]{3.4cm}{Fine-scale Optimization\\ (SIMP method)}}}}
   \rput(4,0.5){\rnode{OutRes}{\psframebox[shadow=true]{\parbox[c]{2.4cm}{\textbf{Output Results}}}}}
   \rput[cl]{0}(10.5,11.5){\scriptsize{COARSE PROBLEM}}
   \psframe[linestyle=dashed](0.8,5)(11.4,11.3)
   \rput[cl]{0}(5.5,2.95){\scriptsize{FINE PROBLEM}}
   \psframe[linestyle=dashed](1.8,1.2)(6.2,2.8)
   \ncline{->}{InData}{MacroDef}
   \ncline{->}{MacroDef}{MacroOpt}
   \ncline{->}{MacroOpt}{DistribDens}
   \ncline{->}{DistribDens}{Conv}
   \ncdiag[angleA=0,armA=2.3cm,angleB=-90,armB=0]{->}{Conv}{UpdateDens}\naput[npos=0.5,labelsep=2pt]{NO}
   \ncline{->}{Conv}{EqTrac}\naput[nrot=0,labelsep=2pt]{YES}
   \ncline{->}{EqTrac}{MicroOpt}
   \ncline{->}{MicroOpt}{OutRes}
   \ncdiag[angleA=90,armA=0.9cm,angleB=0,armB=0]{->}{UpdateDens}{MacroOpt}\nbput[nrot=0,labelsep=2pt]{\(k=k+1\)}
   \end{pspicture}
\end{center}
\caption{General two-level topology optimization procedure}\label{fig2}
\end{figure}

In order to provide an overview of the proposed method (see Figure \ref{fig2}), the main steps of the two-level optimization technique are outlined next. Firstly, a coarse-scale topology optimization problem is solved via the SIMP method with the aim of distributing densities over the whole domain, verifying the usual boundary conditions. Due to the potentially high costs of manufacturing cells with very high (or low) amount of material in relation to completely solid (or void) cells, each optimization is carried out as an iteration into a loop until all densities are in the range of two prescribed thresholds \([\overline \rho_{min},\overline \rho_{max}]\). This updating procedure of densities works in the following way: after each application of the SIMP method, if the density of the element $i$ at the iteration $k$, referred to as $\rho_i^{(k)}$, is outside of the range \([\overline \rho_{min},\overline \rho_{max}]\), it is added to a set of fixed elements which will not be optimized in the following iteration. Specifically, densities $\rho_i^{(k)} \geq \overline \rho_{max}$ are set to $\rho_i^{(k)}=1$ representing solid elements and densities $\rho_i^{(k)} \leq \overline \rho_{min}$ are set to $\rho_i^{(k)}=10^{-3}$ associated to void elements. On the other hand, if an element density is inside the range, its density value is updated and the optimization process is repeated again. Final convergence is reached when densities of all optimized elements in iteration $k$ are in the range \([\overline \rho_{min},\overline \rho_{max}]\) at the end of such iteration.\par

\par

Secondly, an equilibrating tractions method \cite{Ladeveze1996} is applied (see Section \ref{TEquilibrium}). This method uses the converged solution of the coarse problem, whose stress field is discontinuous, to obtain a set of equilibrated tractions on the contour of each coarse cell, which ensures the mechanical continuity among cells. This mechanical continuity represents a relevant improvement to previous works, like \cite{Ferrer2016}, as it will provide structurally viable solutions. Note that if the fine-scale mesh were infinitely small, the final topology in the surroundings of the cell contour will be driven by the local tractions applied to the contour. Since they are equilibrated between cells, the local topology tends to be the same when refining the mesh.\par

Moreover, there is another source of discontinuity between adjacent cells that comes from the different values of relative densities evaluated at each of the cells of the coarse discretization, that is, from the discontinuous density field evaluated in the coarse discretization. This effect depends on the size of the elements of this discretization, but it could be smoothed by using h-adaptive techniques \cite{Munoz2022}.\par

In the last step, each cell is optimized at the fine scale (see Section \ref{SIMPapplication}) under the obtained lateral tractions through the SIMP method again.\par

As in several previous works, e.g. in \cite{Krog1999,Olhoff1998} with microstructure approaches and \cite{Rozvany1996} with the SIMP method, the initial volume fraction $\rho_0$ is adopted as a constant value for each simulation. When the SIMP method is applied, we usually consider $p=3$ to penalize intermedite density values. However, a value of $p=1$ is adopted at the coarse level to obtain a low-definition description of the optimal material distribution, in such a way intermediate densities were not penalized.\par

It should be taken into account that all calculations through the FEM at both levels (coarse and fine) are performed using Cartesian grids \cite{Nadal2013} with four-node square bilinear elements. For implementation simplicity, only domains which can be fitted by Cartesian grids are included. However, the methodology presented in this paper could be extended to other geometries if immersed boundary methodologies, like the Cartesian grid FEM (cgFEM) \cite{Marco2018} or the Finite Cell method \cite{Dauge2015}, were considered.\par

Because of the numerical independency among coarse cells, their optimization might be solved by parallel computing or powerful machine learning strategies to achieve significant improvements in the computational performance with respect to the full-scale approach for similar resolution.

\subsection{The Traction Equilibrating Method}\label{TEquilibrium}
As mentioned in Section \ref{Introd}, the loads for each of the fine-scale topology optimization problem are obtained from the coarse problem solution. The more straightforward way to transfer these loads is by using the stresses obtained from the coarse FEM analysis, as considered in \cite{Ferrer2016}. However, that method presents some difficulties as the FE stresses are discontinuous between elements and because the FE stress field in the element is not equilibrated.\par

\begin{figure}[h! t]
\begin{center}
  \subfigure[Node forces from FEM]{\includegraphics[trim=10pt 0pt 10pt 0pt, clip=false,scale=0.26]{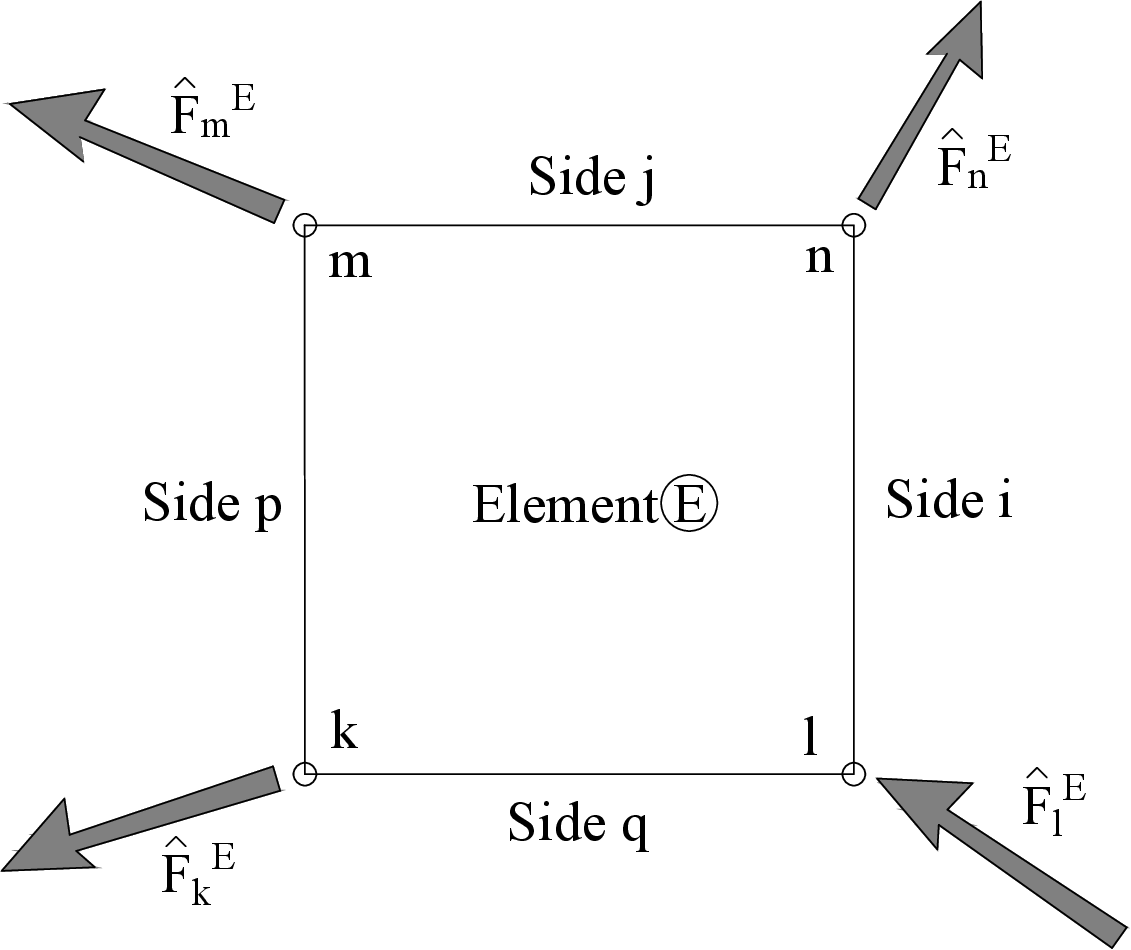}\label{fig4a}}\qquad
  \subfigure[Side nodal forces]{\includegraphics[trim=0pt 10pt 0pt 0pt, clip=false,scale=0.26]{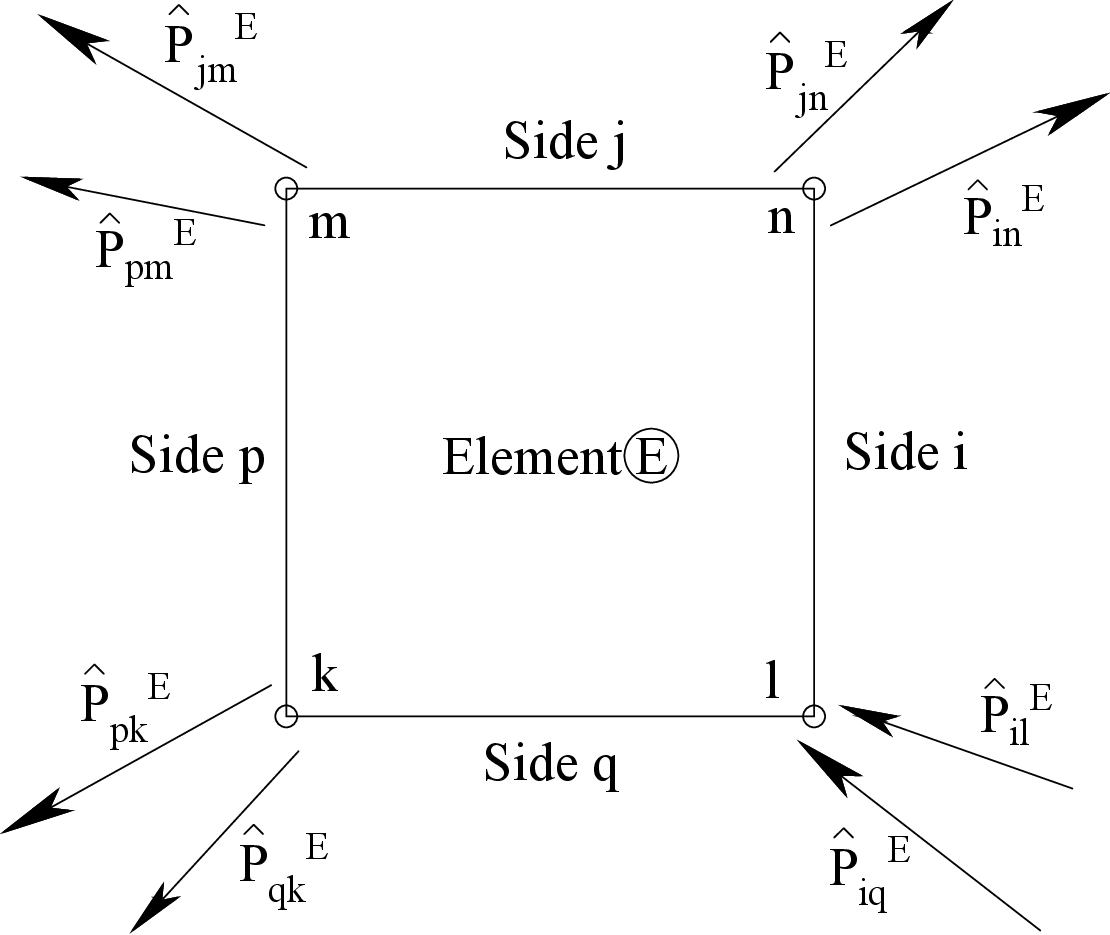}\label{fig4b}}\\
  \subfigure[Traction distribution]{\includegraphics[trim=0pt 0pt 0pt -10pt, clip=false,scale=0.3]{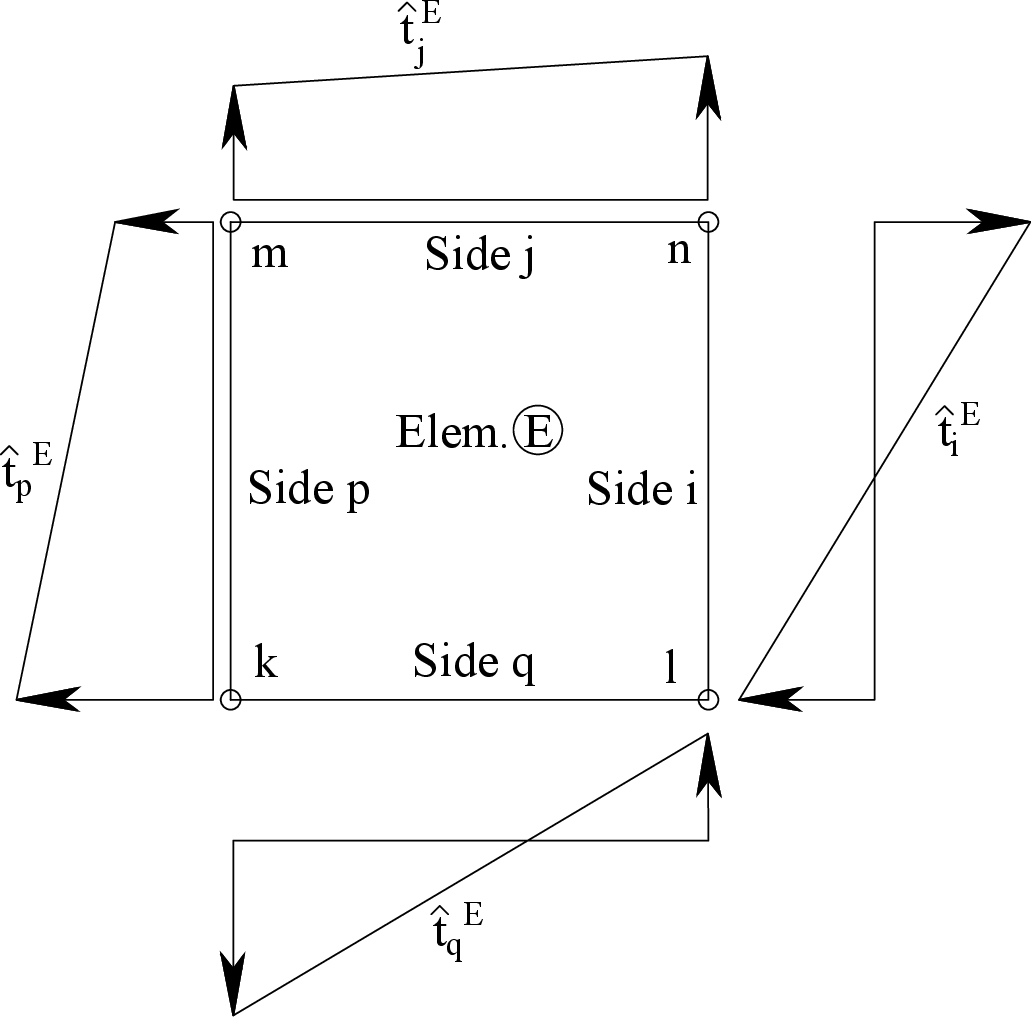}\label{fig4c}}\qquad
  \subfigure[Notation]{\includegraphics[trim=10pt -160pt 10pt 0pt, clip=false,scale=0.22]{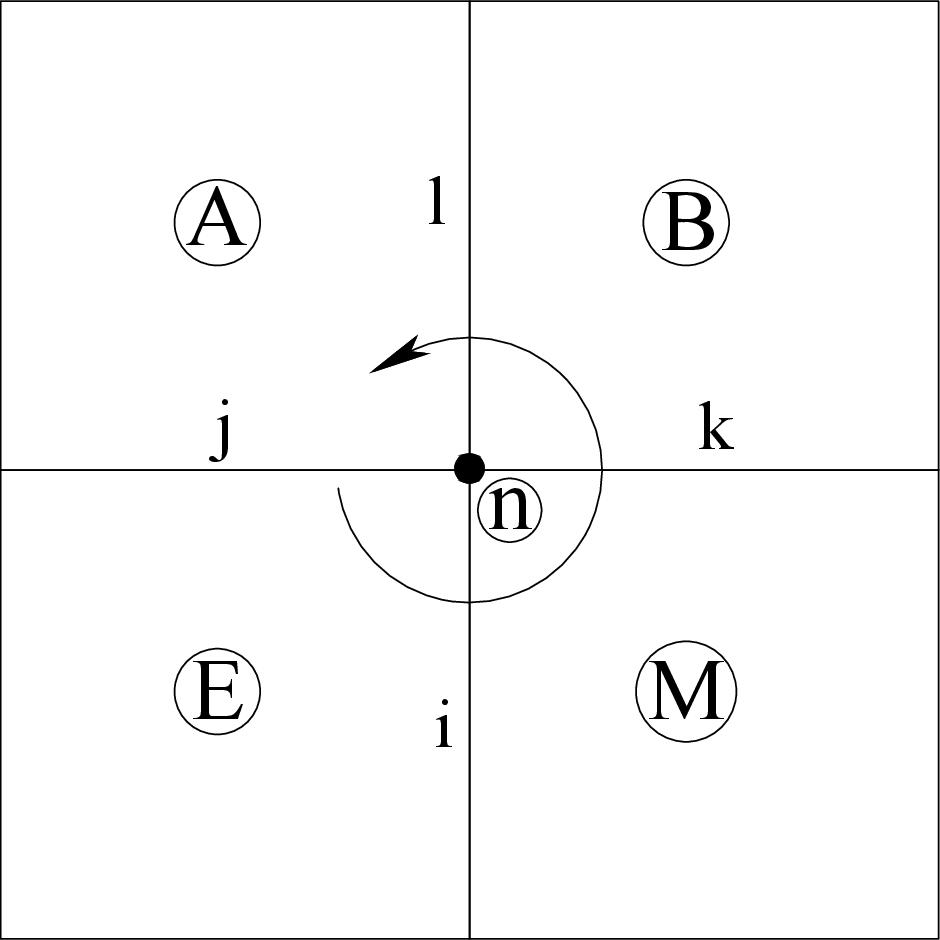}\label{fig4d}}
\end{center}
\caption{Steps and notation in the traction equilibrating method}\label{fig4}
\end{figure}

In this work the traction equilibrating method introduced in \cite{Ladeveze1996} is applied to obtain a continuous traction field along the edges between elements at the coarse level. In addition, we must impose sufficient constraints to prevent the rigid body motion of each coarse element, so that the topology optimization problem at fine level can be solved. Following \cite{Ladeveze1996}, the applied method operates in three main steps (Figure \ref{fig4}): from the FE solution of the coarse problem, the nodal forces $\hat{F}^E_k, \hat{F}^E_l,\hat{F}^E_m, \hat{F}^E_n$ are obtained for each coarse element (Fig. \ref{fig4a}). Each of them (for instance, $\hat{F}^E_n$) is then distributed into two side forces $\hat{P}^E_{in}, \hat{P}^E_{jn}$ (Fig. \ref{fig4b}) in such a way the element satisfies equilibrium with its neighbours. Finally, at each side $i$ of the element a linear traction distribution $\hat{t}_i$, statically equivalent to  $\hat{P}_{il},\ \hat{P}_{im}$, is determined (Fig. \ref{fig4c}). Note that each $\hat{\mathbf{t}}$ will have two components, although, for clarity purposes, only normal stresses have been represented in Figure \ref{fig4c}.\par

Since the calculation details of the side forces $\hat{P}_n$ depend on the number of elements surrounding each specific node into the Cartesian grid, an algorithm was designed to classify each into three main groups: internal nodes, nodes with Dirichlet BC and nodes with Neumann BC.\par

Once the node connectivity is known, the main equations used to solve the side forces $\hat{P}_{in}$ over a coarse-scale element are outlined below (see \cite{Ladeveze1996} for details). Further remarks about the equilibrium of internal nodes surrounding elements which have been turned into void, are described in \ref{Spcases}.

\subsubsection{Internal Nodes}\label{IntNodes}
Nodes without any external force or reaction directly acting on them and surrounded by four elements are included herein. Volume forces are not taken into account throughout this work. For each element, the nodal forces $\hat{F}^e$ are obtained from the FE solution using the following expression:
\begin{equation}\label{eq22}
    \hat{F}^e=K^e u^e,
\end{equation}
where $K^e$ is the stiffness matrix and $u^e$ are the nodal displacements of element $e$.\par

\begin{figure}[h! t]
\begin{center}
  \subfigure[Maxwell polygon at node $n$]{\includegraphics[scale=0.32]{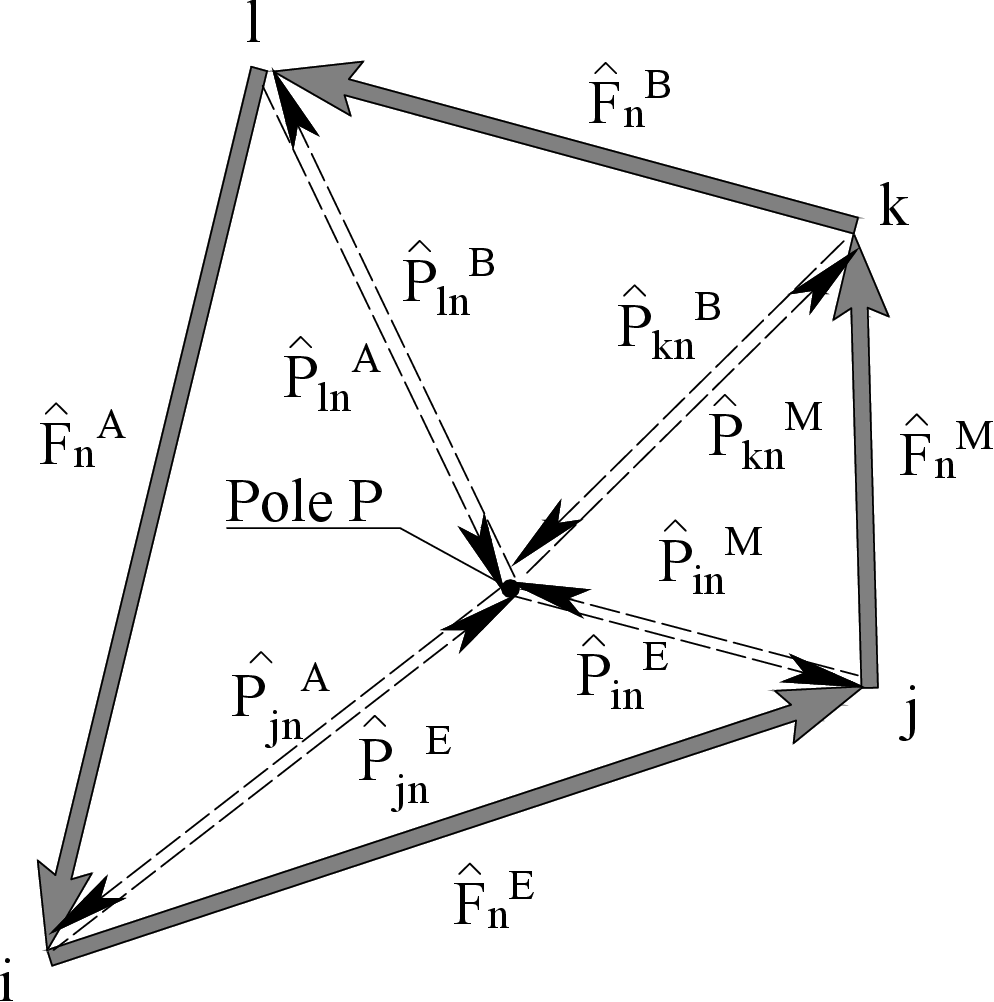}\label{fig51a}}\qquad
  \subfigure[Splitting of the polygon]{\includegraphics[scale=0.32]{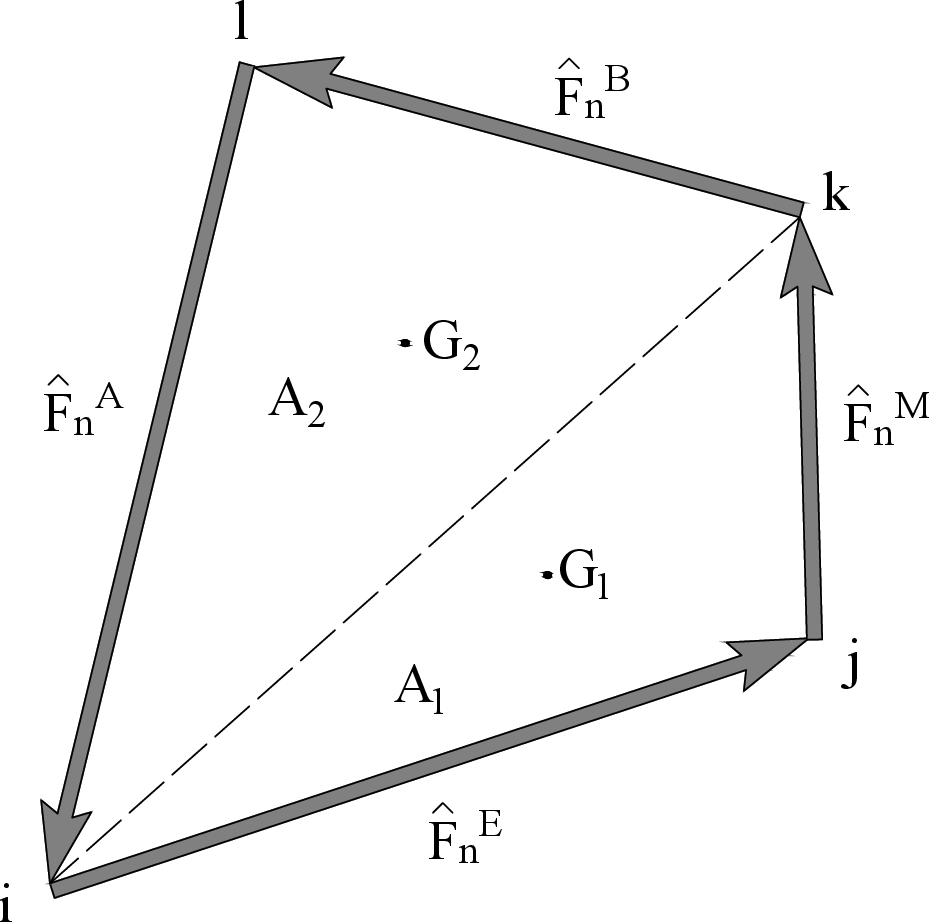}\label{fig51b}}\\
  \subfigure[Strategy to calculate each $G_i$]{\includegraphics[scale=0.33]{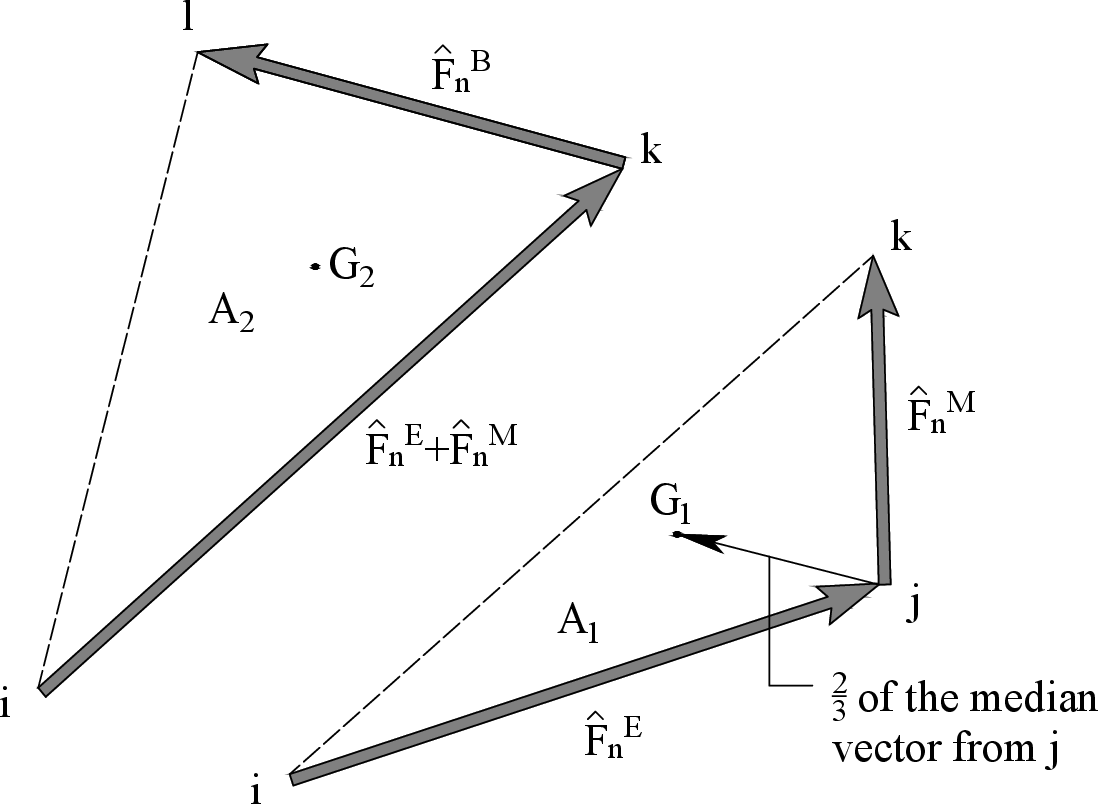}\label{fig51c}}\qquad
  \subfigure[Example of intersection between opposite sides]{\includegraphics[scale=0.32]{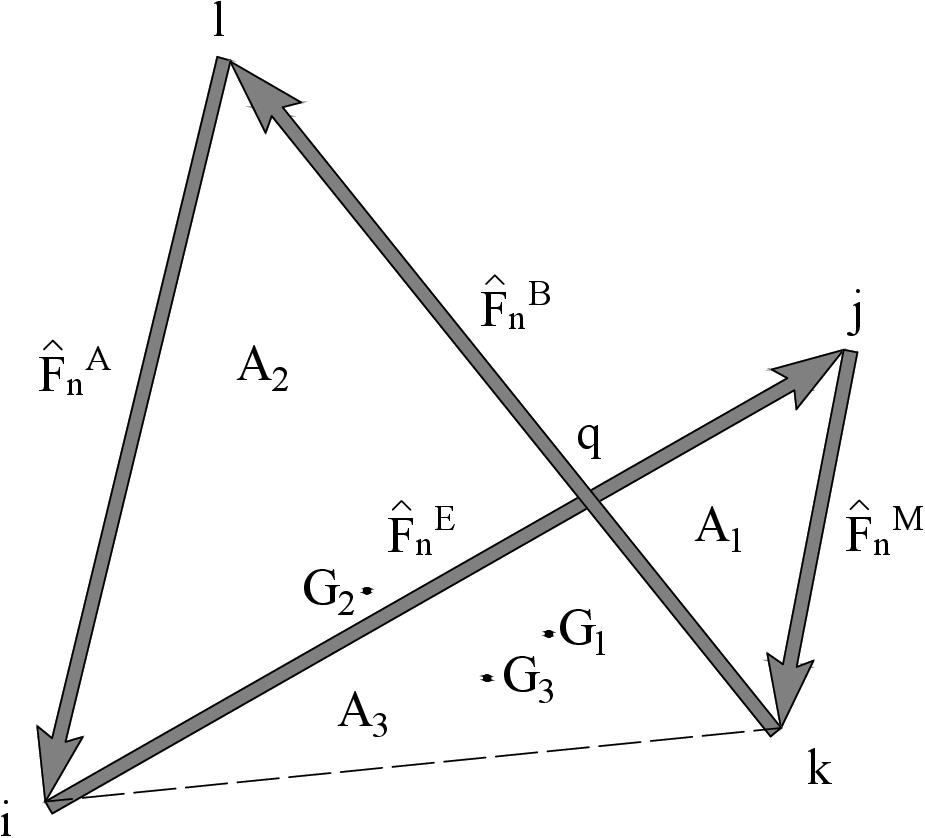}\label{fig51d}}\\
\end{center}
\caption{Stages to calculate the center of masses of Maxwell polygon}\label{fig51}
\end{figure}

Consider elements $E,M,B,A$ connected to node $n$ as shown in Figure \ref{fig4d}. As nodes are in equilibrium, if we use a Maxwell diagram to represent the nodal forces from these elements ($\hat{F}_n^E, \hat{F}_n^M, \hat{F}_n^B, \hat{F}_n^A$) acting on $n$, we obtain a closed quadrilateral polygon (see Figure \ref{fig51a}). In addition, the action-reaction principle, $\hat{P}_{in}^E=-\hat{P}_{in}^M$, and the equilibrium in each corner, $\hat{P}_{in}^E+\hat{P}_{jn}^E=\hat{F}_n^E$ (Fig. \ref{fig4b}) are kept for the side nodal forces, as graphically shown in Fig. \ref{fig51a}, where point P is called \emph{pole}. The force splitting $\hat{P}_{in}^E+\hat{P}_{jn}^E=\hat{F}_n^E$ is not unique, hence, we can select one of the possible solutions simply selecting the location of the \emph{pole} P. Although this location could be arbitrarily selected, following \cite{Ladeveze1996}, in order to avoid excessively large side nodal forces and to avoid a large computational cost, the \emph{pole} has been chosen at the geometrical mass centre of the polygon as
\begin{equation}\label{eq23}
    \hat{P}_{in}^E=r_G,
\end{equation}
where the origin of coordinates of the polygon has been taken at vertex $i$ in figure \ref{fig51a}, which includes the Maxwell polygon for the sake of completeness. In any case, other approaches can be followed as described in \cite{Ladeveze1996}.\par

Rearranging the mentioned equations in matrix form, the following system is derived:
\begin{equation}\label{eq24}
    \begin{bmatrix}
      1 & 1 & 0 & 0 & 0 & 0 & 0 & 0 & 1 \\
      0 & 0 & 1 & 1 & 0 & 0 & 0 & 0 & 0 \\
      0 & 0 & 0 & 0 & 1 & 1 & 0 & 0 & 0 \\
      0 & 0 & 0 & 0 & 0 & 0 & 1 & 1 & 0 \\
      1 & 0 & 0 & 0 & 0 & 0 & 0 & 1 & 0 \\
      0 & 1 & 1 & 0 & 0 & 0 & 0 & 0 & 0 \\
      0 & 0 & 0 & 1 & 1 & 0 & 0 & 0 & 0 \\
      0 & 0 & 0 & 0 & 0 & 1 & 1 & 0 & 0 \\
      1 & 0 & 0 & 0 & 0 & 0 & 0 & 0 & 0 \\
    \end{bmatrix}\begin{Bmatrix}
                      \hat{P}_{in}^E\\
                      \hat{P}_{jn}^E\\
                      \hat{P}_{jn}^A\\
                      \hat{P}_{ln}^A\\
                      \hat{P}_{ln}^B\\
                      \hat{P}_{kn}^B\\
                      \hat{P}_{kn}^M\\
                      \hat{P}_{in}^M\\
                      \lambda\\
                    \end{Bmatrix}=\begin{Bmatrix}
                      \hat{F}_{n}^E\\
                      \hat{F}_{n}^A\\
                      \hat{F}_{n}^B\\
                      \hat{F}_{n}^M\\
                      0\\
                      0\\
                      0\\
                      0\\
                      r_G\\
                    \end{Bmatrix},
\end{equation}
where $\lambda$ is a force residual parameter introduced to make the system in \eqref{eq24} well-posed and will end up to be zero. It should be noted that each of the forces $\hat{P}_{in},\ \hat{F}_n$ and $\lambda$ have two components (x and y directions). Thus, the system \eqref{eq24} is sized $18 \times 18$ and its solution will provide the eight side nodal forces $\hat{P}^E$ for each element (two per side).\par

A simple strategy has been adopted to obtain the geometrical mass centre of the Maxwell polygon. Firstly, the quadrilateral is divided into two triangles along its diagonal $i-k$ (Figure \ref{fig51b}) and secondly, the areas and center of masses of both triangles are obtained by using two edge vectors taken in an anticlockwise sense (Figure \ref{fig51c}), obtaining (e.g. for $ijk$) their properties from:
\begin{subequations}\label{eq24.1}
\begin{align}
     & A_1 = \frac{1}{2} (\hat{F}_n^E \times \hat{F}_n^M)_z \label{eq24.1a}\\
     & r_{G_1} = \hat{F}_n^E + \frac{2}{3}\mu,           \label{eq24.1b}
\end{align}
\end{subequations}
where $\mu$ is the median vector from vertex $j$ and $(\ )_z$ indicates the third component of the cross product.\par

Equations \eqref{eq24.1} can be adapted to triangle $ikl$. The final geometrical mass centre of the quadrilateral is obtained as
\begin{equation}\label{eq24.2}
    r_{G}=\frac{r_{G_1} A_1+r_{G_2} A_2}{A_1+A_2}.
\end{equation}
Note that the possible \emph{concavity} in one vertex of the quadrilateral does not modify equation \eqref{eq24.2}, even if the sense of the cross product \eqref{eq24.1a} were inverted. Nevertheless, if this \emph{concavity} leads to intersect two opposite sides of the polygon (see Fig. \ref{fig51d} as an example), \eqref{eq24.2} should be replaced by
\begin{equation}\label{eq24.3}
    r_{G}=\frac{r_{G_1} |A_1|+r_{G_2} |A_2|-2r_{G_3} |A_3|}{|A_1|+|A_2|-2|A_3|}
\end{equation}
in order to remove overlapping areas in triangle $ikq$ and the subsequent inconsistency of equilibrium.

The following equation is used in the FEM to evaluate equivalent nodal forces $\hat{P}^E_{in}$ from tractions $\hat{t}^E_i$ applied on element sides:
\begin{equation}\label{eq24.4}
    \int_i \phi_n^E \hat{t}^E_i dS = \int_i \phi_n^E (\phi_n^E \hat{t}^E_{in} + \phi_m^E \hat{t}^E_{im}) dS = \hat{P}^E_{in},
\end{equation}
\noindent where $\phi_n^E, \phi_m^E$ represent the standard shape functions along side $i$. In our case, $\hat{P}^E_{in}$ is obtained from \eqref{eq24}, hence, from \eqref{eq24.4} we can evaluate the nodal values of $\hat{t}_{in}^E,\ \hat{t}_{im}^E$ that define the linear tractions on side $i$ solving the following system of equations:
\begin{equation}\label{eq24.5}
    \begin{bmatrix}
        \int_i (\phi_n^E)^2 dS & \int_i (\phi_n^E \phi_m^E) dS\\
        \int_i (\phi_m^E \phi_n^E) dS & \int_i (\phi_m^E)^2 dS\\
    \end{bmatrix}\begin{Bmatrix}
                    \hat{t}_{in}^E\\
                    \hat{t}_{im}^E
                \end{Bmatrix} = \begin{Bmatrix}
                                    \hat{P}^E_{in}\\
                                    \hat{P}^E_{im}
                                \end{Bmatrix},
\end{equation}
The resulting values of system \eqref{eq24.5} will be used afterwards in the fine-scale topology optimization.\par

Sections \ref{DNodes} and \ref{NNodes} will outline the main equations used for nodes under BC. Further details can be found in \cite{Ladeveze1996}.

\subsubsection{Nodes with Dirichlet BC}\label{DNodes}
In the case of 2D Cartesian grids these nodes will be surrounded either by one or by two elements.
\begin{description}
  \item[Standard case.] The first case is defined by two elements surrounding a node on $\Gamma_D$. In this case, the reaction force will be divided in two components $R_n^E,R_n^M$ that are calculated together with the interface forces $\hat{P}_{in}^E,\ \hat{P}_{in}^M$. Thus, the Maxwell diagram is a triangle and the pole is located at the center of masses through eq. \eqref{eq23}, which is equivalent in this case to
      \begin{equation}\label{eq25}
        \hat{P}_{in}^E={R}_n^E-{R}_n^M.
      \end{equation}\par

      Rearranging equations in matrix form, the system that represents the equilibrium and the action-reaction principle is written as
      \begin{equation}\label{eq26}
        \begin{bmatrix}
          1 & 1 & 0 & 0 & -1\\
          0 & 0 & 1 & 1 & 0\\
          0 & 1 & 0 & 1 & 0\\
          1 & 0 & 1 & 0 & -1\\
          -1 & 1 & 0 & -1 & 0\\
        \end{bmatrix}\begin{Bmatrix}
                      \hat{P}_{in}^E\\
                      R_n^E\\
                      \hat{P}_{in}^M\\
                      R_n^M\\
                      \lambda
        \end{Bmatrix}=\begin{Bmatrix}
                        \hat{F}_n^E\\
                        \hat{F}_n^M\\
                      R_n\\
                      0\\
                      0\\
        \end{Bmatrix}.
      \end{equation}
    \item[Outer corner node.] Assuming a reaction $R_n$ is applied on the node, its module is equal to the node force, but with opposite sign. This means that there is no force left to add, as this node will be equilibrated. Therefore the forces applied on this node are obtained straightaway.
\end{description}

\subsubsection{Nodes with Neumann BC}\label{NNodes}
The analysis of nodes under Neumann BC is quite similar to that in section \ref{DNodes}. However, the external forces are known since our developed code properly distributes the tractions on nodes in $\Gamma_N$, previous to the FEM analysis of the optimization process. The following three cases are studied:

\begin{description}
  \item[Standard case] The first case is defined as a boundary node surrounded by two elements under Neumann BC, the external forces $T_n^E,T_n^M$ applied in each element connected with the same node are calculated previously to the FEM optimization analysis. Now, the Maxwell diagram is a quadrilateral with the \emph{pole} located at one vertex (see Fig. 6 in \cite{Ladeveze1996}). The corresponding system of equations takes the form:
      \begin{equation}\label{eq27}
        \begin{bmatrix}
          1 & 0 & 1\\
          0 & 1 & 1\\
          1 & 1 & 0\\
        \end{bmatrix}\begin{Bmatrix}
                      \hat{P}_{in}^E\\
                      \hat{P}_{in}^M\\
                      \lambda
        \end{Bmatrix}=\begin{Bmatrix}
                        \hat{F}_n^E-T_n^E\\
                        \hat{F}_n^M-T_n^M\\
                        0\\
        \end{Bmatrix}.
      \end{equation}
      Obviously, when no external forces are applied, Eq. \eqref{eq27} becomes
      \begin{equation}\label{eq28}
        \begin{bmatrix}
          1 & 0 & 1\\
          0 & 1 & 1\\
          1 & 1 & 0\\
        \end{bmatrix}\begin{Bmatrix}
                      \hat{P}_{in}^E\\
                      \hat{P}_{in}^M\\
                      \lambda
        \end{Bmatrix}=\begin{Bmatrix}
                        \hat{F}_n^E\\
                        \hat{F}_n^M\\
                        0\\
        \end{Bmatrix}.
      \end{equation}
    \item[Outer corner node.] This case is as simple as the one with Dirichlet BC, but here instead of using the reaction force, the known external force $T_n$ is applied. Once again the resulting force is obtained straightaway since the two forces are equilibrated. When no forces are applied, the node is treated as a free node and will only be affected by the corner node force $\hat{F}_n^E$.
    \item[Reentrant corner node.] Let us consider a reentrant corner formed by elements A, E and M in Figure \ref{fig4d} with their known external forces $T_n^A, T_n^E,T_n^M$. In this case, the Maxwell diagram becomes a triangle with the \emph{pole} located at one vertex. Hence, the system of equations of equilibrium can be written as:
      \begin{equation}\label{eq281}
        \begin{bmatrix}
          1 & 0 & 0 & 0 & 1\\
          0 & 1 & 0 & 0 & 0\\
          0 & 0 & 1 & 1 & 0\\
          0 & 1 & 1 & 0 & 1\\
          1 & 0 & 0 & 1 & 0\\
        \end{bmatrix}\begin{Bmatrix}
                      \hat{P}_{jn}^A\\
                      \hat{P}_{in}^M\\
                      \hat{P}_{in}^E\\
                      \hat{P}_{jn}^E\\
                      \lambda
        \end{Bmatrix}=\begin{Bmatrix}
                        \hat{F}_n^A-T_n^A\\
                        \hat{F}_n^M-T_n^M\\
                        \hat{F}_n^E-T_n^E\\
                        0\\
                        0\\
        \end{Bmatrix}.
      \end{equation}
      Thus, when no external forces are applied, Eq. \eqref{eq281} is reduced to
      \begin{equation}\label{eq282}
        \begin{bmatrix}
          1 & 0 & 0 & 0 & 1\\
          0 & 1 & 0 & 0 & 0\\
          0 & 0 & 1 & 1 & 0\\
          0 & 1 & 1 & 0 & 1\\
          1 & 0 & 0 & 1 & 0\\
        \end{bmatrix}\begin{Bmatrix}
                      \hat{P}_{jn}^A\\
                      \hat{P}_{in}^M\\
                      \hat{P}_{in}^E\\
                      \hat{P}_{jn}^E\\
                      \lambda
        \end{Bmatrix}=\begin{Bmatrix}
                        \hat{F}_n^A\\
                        \hat{F}_n^M\\
                        \hat{F}_n^E\\
                        0\\
                        0\\
        \end{Bmatrix}.
      \end{equation}
\end{description}

\subsection{Application of the SIMP Method at the fine scale}\label{SIMPapplication}
Once the equilibrated tractions at the coarse scale have been obtained using the procedure previously described, we will solve another topology optimization problem in each element of this coarse scale, considering these equilibrated tractions as Neumann BC and a fine-scale discretization.\par

Since every fine-scale problem is defined into a square domain, the SIMP method presented in \cite{Sigmund2001} was used for the sake of simplicity, but slightly modified with the projection method proposed in \cite{Wang2011}, to solve the topology optimization problems at the fine scale. A flowchart of the proposed procedure is shown in Figure \ref{fig6}. Although the standard SIMP method is well-known, several comments are appropriate to clarify this flowchart:

\begin{figure}[h! t]
\begin{center}
 \begin{pspicture}(0,0)(9,12)
   \rput(3.5,11.5){\rnode{MACSS}{\psframebox[shadow=true]{\parbox[c]{3cm}{Coarse-scale solution}}}}
   \rput(3.5,10.5){\rnode{IniDensity}{\psframebox{\parbox[c]{2.5cm}{Initial density $\rho_e^{(0)}$}}}}
   \rput(3.5,9.5){\rnode{FEM}{\psframebox{\parbox[c]{2.5cm}{FEM solution $\mathbf{u}^{(k)}$}}}}
   \rput(3.5,8.5){\rnode{Sensitivity}{\psframebox{\parbox[c]{4.4cm}{Sensitivity evaluation $\left[\frac{\partial c(\rho)}{\partial \rho}\right]_{\mathbf{u}^{(k)}}$}}}}
   \rput(3.5,7.5){\rnode{Filter}{\psframebox{\parbox[c]{4cm}{Filtering, eqs. (5),(6) in \cite{Sigmund2001}}}}}
   \rput(3.5,6.6){\rnode{UpdDensity}{\psframebox{\parbox[c]{3.2cm}{Density updating \eqref{eq20}}}}}
   \rput(3.5,5.5){\rnode{Dproj}{\psframebox{\parbox[c]{4cm}{Density projection, eq. \eqref{eq28.4}}}}}
   \rput(3.5,4.3){\rnode{Mnd}{\psframebox{\parbox[c]{3cm}{Calculation $M_{nd}^{(k)}$ \eqref{eq29}}}}}
   \rput(3.5,3.2){\rnode{CheckMnd}{\psdiabox[framesep=0.05]{\parbox[c]{1.9cm}{\footnotesize{$M_{nd}^{(k)}>M_{nd,min}$?}}}}}
   \rput(7.7,3.2){\rnode{UpdBeta}{\psframebox{\parbox[c]{2.8cm}{\small{Updating}\\ \footnotesize{$\rho_e^{(k+1)} = \Tilde{\rho}_e^{(k+1)}$},\\ \footnotesize{$\beta^{(k+1)} = \min\{2\beta^{(k)},\beta_{max}\}$} }}}}
   \rput(3.5,1.8){\rnode{Conv}{\psdiabox{\parbox[c]{2.5cm}{\footnotesize{$\max |\rho_e^{(k+1)}-\rho_e^{(k)}| < \varepsilon?$}}}}}
   \rput(3.5,0.5){\rnode{Results}{\psframebox[shadow=true]{\parbox[c]{2cm}{Output results}}}}
   \ncline{->}{MACSS}{IniDensity}
   \ncline{->}{IniDensity}{FEM}
   \ncline{->}{FEM}{Sensitivity}
   \ncline{->}{Sensitivity}{Filter}
   \ncline{->}{Filter}{UpdDensity}
   \ncline{->}{UpdDensity}{Dproj}\naput[labelsep=2pt]{\footnotesize{$\rho_e^{(k+1)}$}}
   \ncline{->}{Dproj}{Mnd}\naput[labelsep=2pt]{\footnotesize{$\Tilde{\rho}_e^{(k+1)}$}}
   \ncline{->}{Mnd}{CheckMnd}
   \ncline{->}{CheckMnd}{Conv}\naput[labelsep=2pt]{YES}
   \ncline{->}{CheckMnd}{UpdBeta}\naput[labelsep=2pt]{NO}
   \ncdiag[angleA=270,armA=0cm,angleB=0,armB=1.5cm]{->}{UpdBeta}{Conv}
   \ncline{->}{Conv}{Results}\naput[labelsep=2pt]{YES}
   \ncdiag[angleA=180,armA=0.3cm,angleB=180,armB=1.6]{->}{Conv}{FEM}\nbput[labelsep=2pt,nrot=90]{NO}\naput[labelsep=2pt,nrot=90]{\(k=k+1\)}
\end{pspicture}
\end{center}
\caption{Application of the SIMP method at the fine scale.}\label{fig6}
\end{figure}
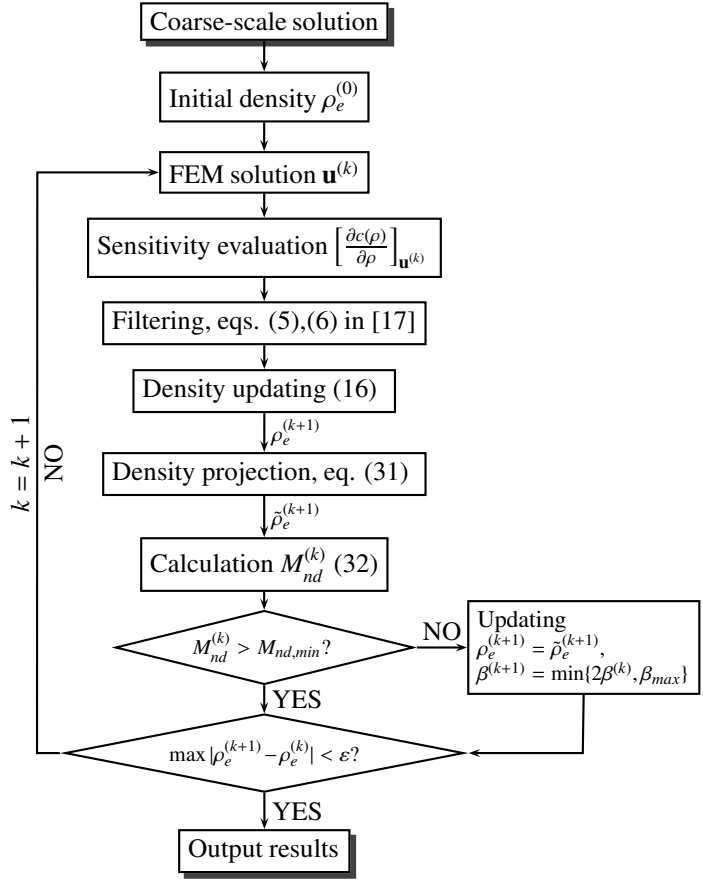

\begin{enumerate}[a.]
    \item As in the general procedure shown in Figure \ref{fig2}, the density of each element $e$ in the fine scale at iteration $k$, is referred to as $\rho_e^{(k)}$.
    \item In order to remove rigid body motions, simply support conditions have been adopted for the square domains that define the fine scale. Since the lateral tractions applied on each cell are equilibrated, the support reactions vanish and do not affect to the internal behaviour at the fine scale.
    \item As the basic 99-line SIMP code introduced in \cite{Sigmund2001} produced undesired gray areas in some coarse-scale square cells, the exponential density projection \eqref{eq28.4} proposed by \cite{Wang2011} with a density threshold of $\mu = 0.5$ was adopted to improve the convergence of the process:
    \begin{equation}\label{eq28.4}
        \Tilde{\rho}_e= \begin{cases}
                            \mu[e^{-\beta(1-\rho_e/\mu)}-(1-\rho_e/\mu)e^{-\beta}] & \hspace{30pt} 0 \leq \rho_i \leq \mu \\
                            (1-\mu)[1-e^{-\beta(\rho_e-\mu)/(1-\mu)}+(\rho_e-\mu)e^{-\beta}/(1-\mu)]+\mu & \hspace{30pt} \mu < \rho_i \leq 1
                        \end{cases}.
    \end{equation}
    \item Focused on controlling the convergence more accurately, the projection of densities and the updating of the projection parameter $\beta^{(k)}$ are carried out every 2 iterations up to convergence. Indeed, $\beta^{(k)}$ is forced to remain below the value of the input parameter $\beta_{max}$. Otherwise, disjointed solutions with material regions separated by large void regions may appear.
    \item A way to measure the level of sharpness obtained in the fine-scale solution, the following parameter (see \cite{Sigmund2007}) is calculated over each iteration:
     \begin{equation}\label{eq29}
       M_{nd}^{(k)}=\sum_{e=1}^n \frac{4\rho_e^{(k)}(1-\rho_e^{(k)})}{n} 100\% \in [0,100],
     \end{equation}
     where $n$ is the total number of elements in the fine-scale FE mesh. As can be expected, very high or very low values of densities $\rho_e^{(k)}$ produce low values of $M_{nd}^{(k)}$, while intermediate values of $\rho_e^{(k)}$ produce high values of $M_{nd}^{(k)}$. Therefore, the parameter $M_{nd}^{(k)}$ is useful to identify fine-scale solutions with too many elements having intermediate densities. Thus, if $M_{nd}^{(k)}$ is greater than certain value of the input parameter $M_{nd,min}$, the associated gray level is considered excessive and the corresponding density projection is carried out. Otherwise, the solution is considered to be sharp enough and the projection does not take place.
     \item Although in the coarse scale we considered $p=1$ because we wanted to obtain a coarse representation of the density distribution, we use $p=3$ in the fine-scale topology optimization process to sharpen the final solution penalizing intermediate density values.
\end{enumerate}

\section{Numerical Results}\label{results}
In the present section, the application of the proposed two-level topology optimization method is studied considering two numerical examples under plane stress. The first one is based on the classic cantilever beam problem whereas the second one is a cantilevered L-shape domain problem that produces a singular stress field. Bilinear quadrilateral elements with Young Modulus $E=1000$ MPa and poison ratio $\nu=0.3$ were used in both problems.\par

\subsection{Example 1. Cantilever beam under pure shear loading}
\begin{figure}[h! t]
\begin{center}
    \includegraphics[trim=0pt 40pt 0pt 30pt, clip=true,scale=0.4]{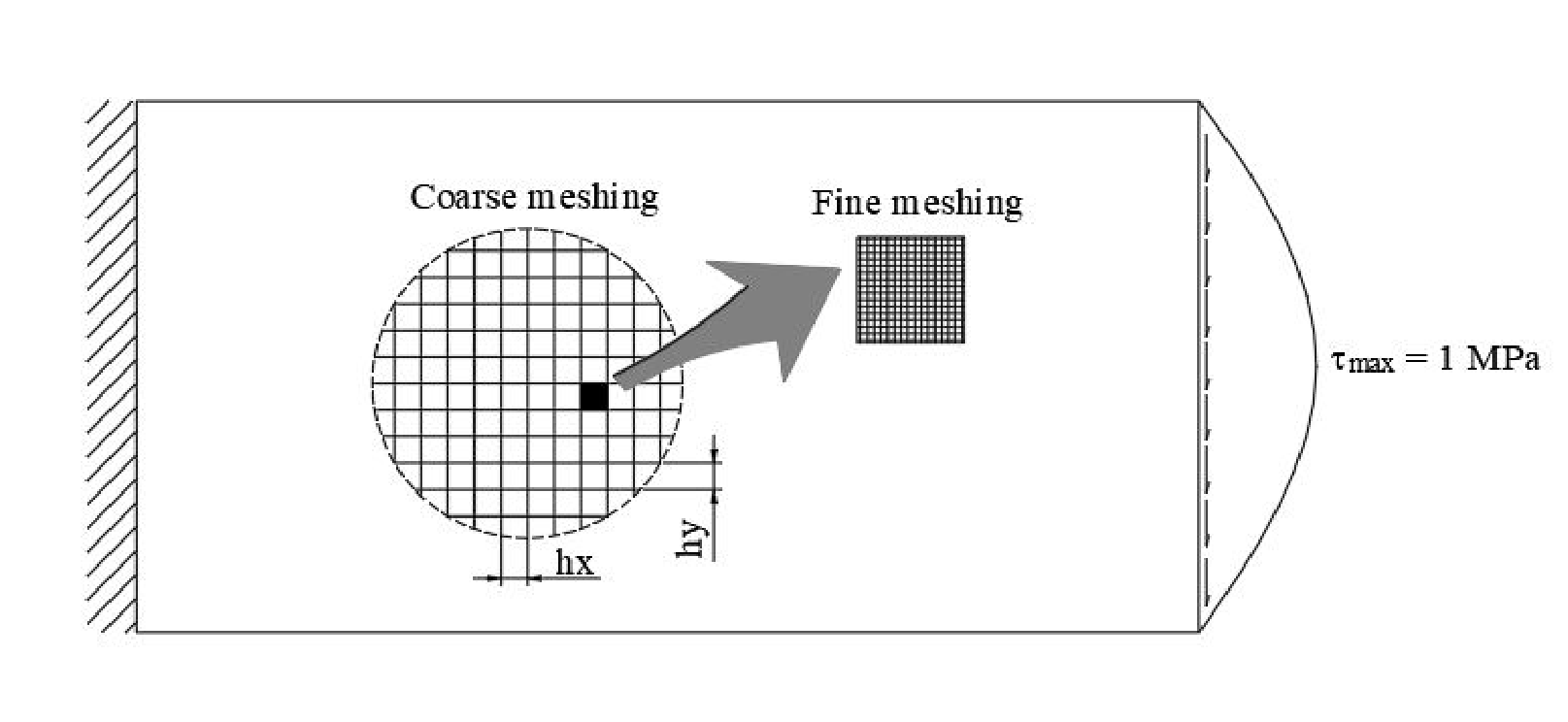}
\end{center}
\caption{Example 1, external loads and meshing.}\label{fig7}
\end{figure}

In the first example the applied load  at the free end of the beam is defined as a parabolic shear stress distribution with $\tau_{max} = 1$ MPa, as shown in Figure \ref{fig7}. Beam dimensions are $2 \times 1$ m and the prescribed volume fraction is $\rho_0 = 0.5$.  Other input parameters for this problem are outlined in Table \ref{tab1}.\par

\begin{table}[h! t]
\begin{center}
\begin{tabular}{lll}
\hline
Coarse scale & Fine scale & Projection\\
\hline
$32 \times 16$ elems. & $32 \times 32$ elems. & $\beta_0 = 1$\\
$p=1$  & $p=3$ & $\beta_{max} = 2$\\
$r_{min} = 1.5$  & $r_{min} = 1.3$ & $\mu = 0.5$\\
$\varepsilon = 0.03$ & $\varepsilon = 0.01$ & $M_{nd,min} = 50\%$\\
\hline
\end{tabular}
\end{center}
\caption{Input parameters for example 1.}\label{tab1}
\end{table}

\begin{figure}[h! t]
\begin{center}
  \subfigure[Stage 1]{\includegraphics[trim=50pt 80pt 40pt 50pt, clip=true, scale=0.46]{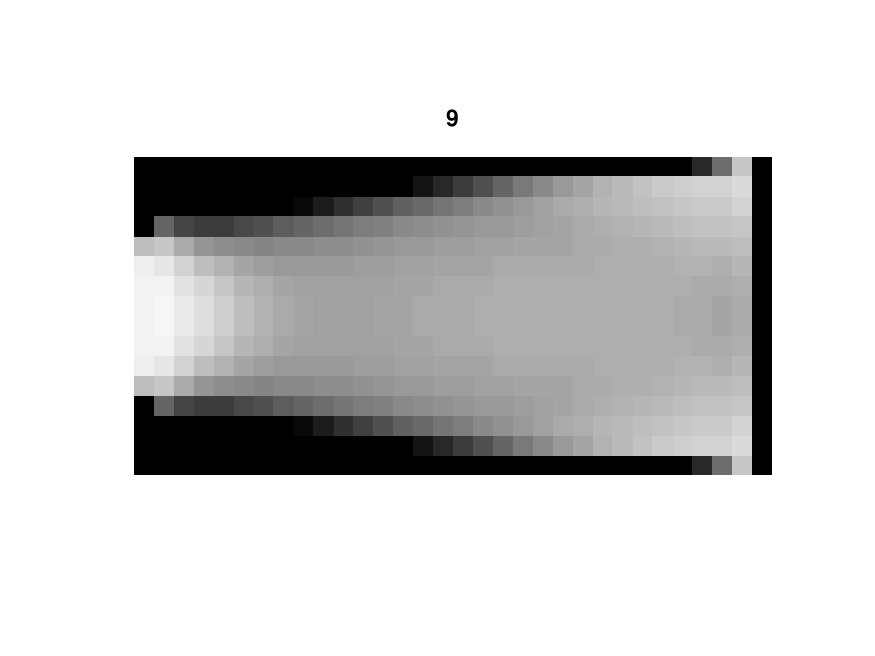}\label{fig8a}}\hskip 3pt
  \subfigure[Stage 3]{\includegraphics[trim=50pt 80pt 40pt 50pt, clip=true, scale=0.46]{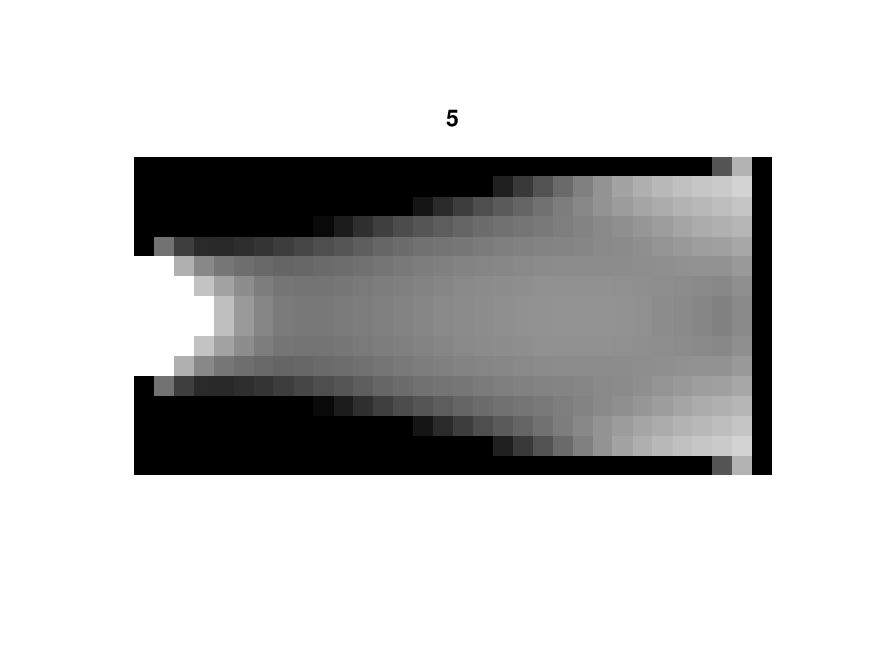}\label{fig8b}}\hskip 3pt
  \subfigure[Stage 5]{\includegraphics[trim=50pt 80pt 40pt 50pt, clip=true, scale=0.46]{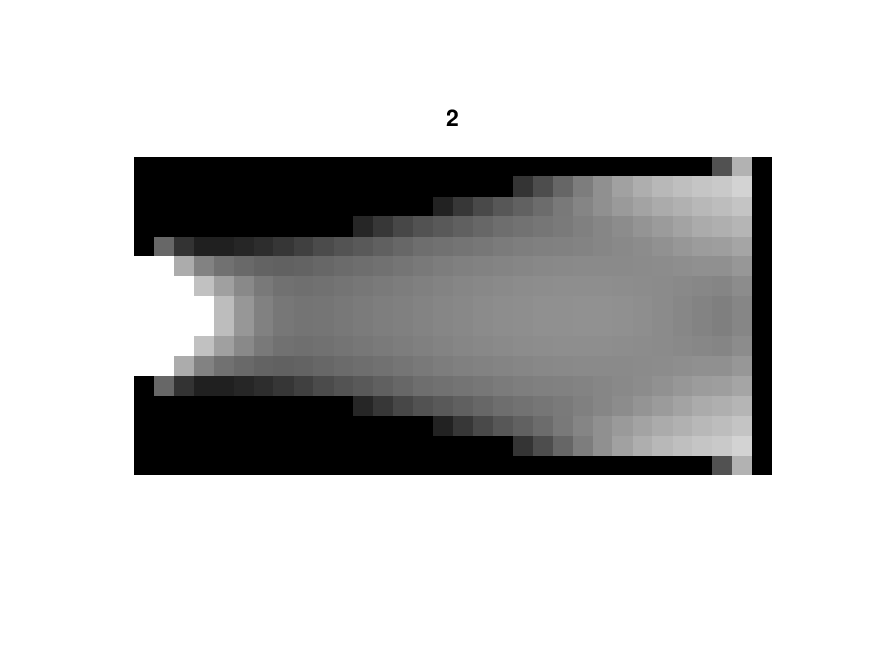}\label{fig8c}}\\
   \subfigure{\includegraphics[trim=0pt 40pt 0pt 400pt, clip=true, scale=0.35]{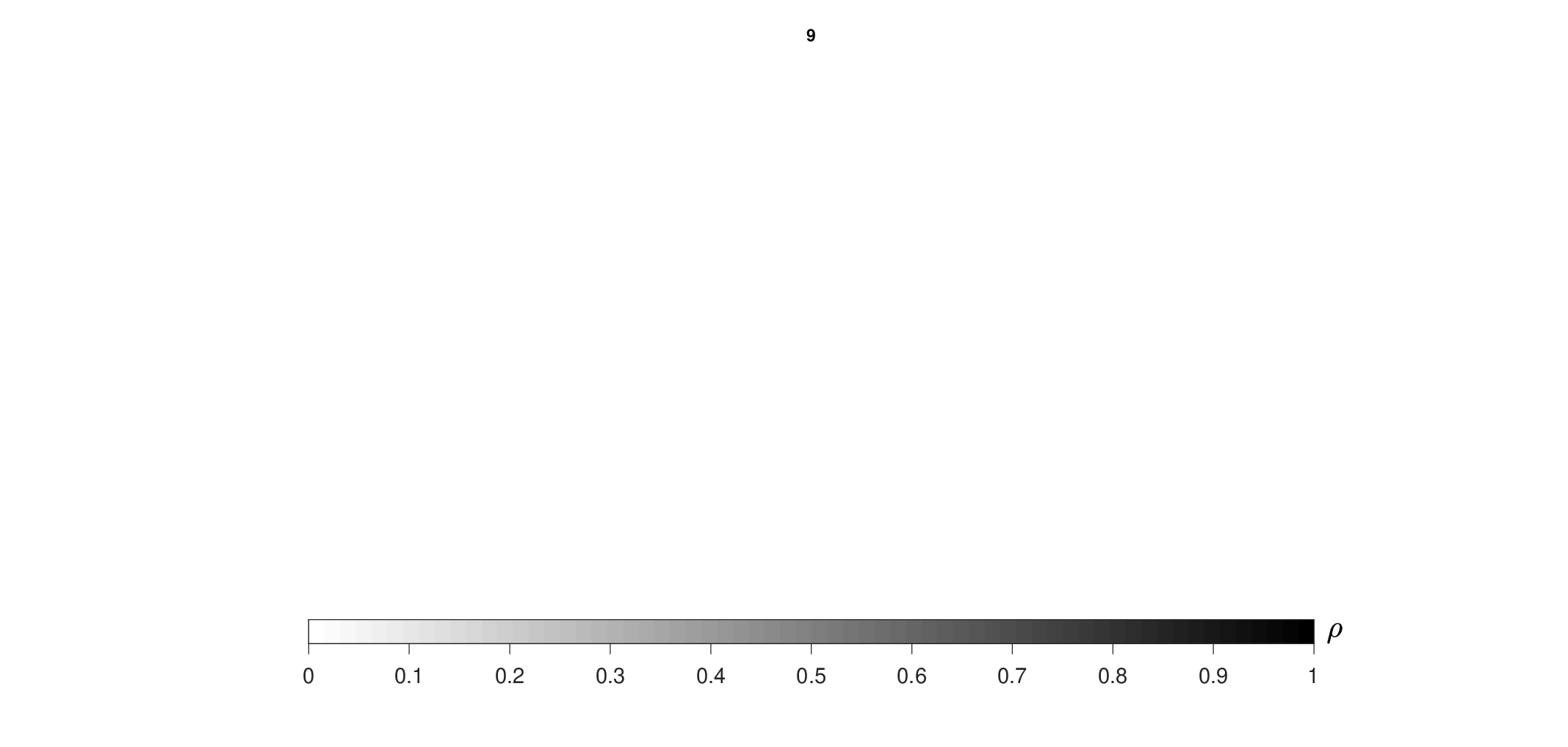} \notag}
\end{center}
\caption{Example 1, density distribution of the coarse problem with \([\overline \rho_{min},\overline \rho_{max}]=[0.12,0.88]\).}\label{fig8}
\end{figure}

A first test with \([\overline \rho_{min},\overline \rho_{max}]=[0.12,0.88]\) is considered to show the performance of the proposed method. The coarse problem described in Figure \ref{fig2} required five stages up to convergence, some of them illustrated in Figure \ref{fig8}, where the number over the graph stands for the iterations up to convergence in the application of the SIMP method into each stage. \par

As expected, the material is accumulated on regions with high values of normal stresses due to the bending moment (darker elements on top and bottom regions), dark elements of high density are also obtained at the right-hand side, where the external load is applied. The central region of the domain appears light gray because of the low stress values produced by the shear force. Regarding the density distribution, note that the central left-hand region progressively tends to be a void region because of its low stress values.\par

\begin{figure}[h! t]
\begin{center}
    \subfigure[Final solution]{\includegraphics[trim=130pt 45pt 50pt 0pt, clip=true,scale=0.32]{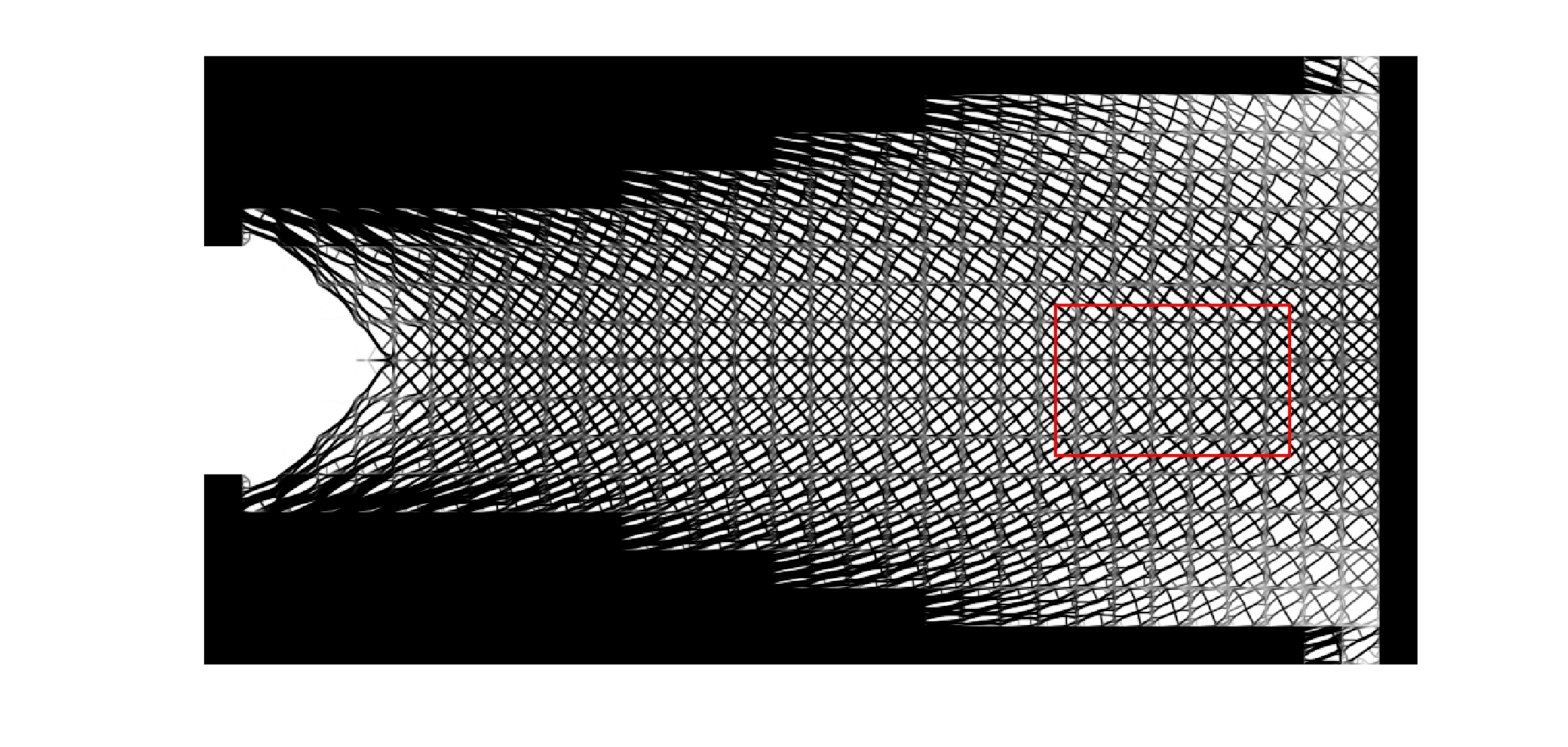}\label{fig9a}}\hspace{3pt}
    \subfigure[Zoom of the rectangular area]{\includegraphics[trim=140pt 50pt 120pt 0pt, clip=true, width = 180pt, height = 140pt]{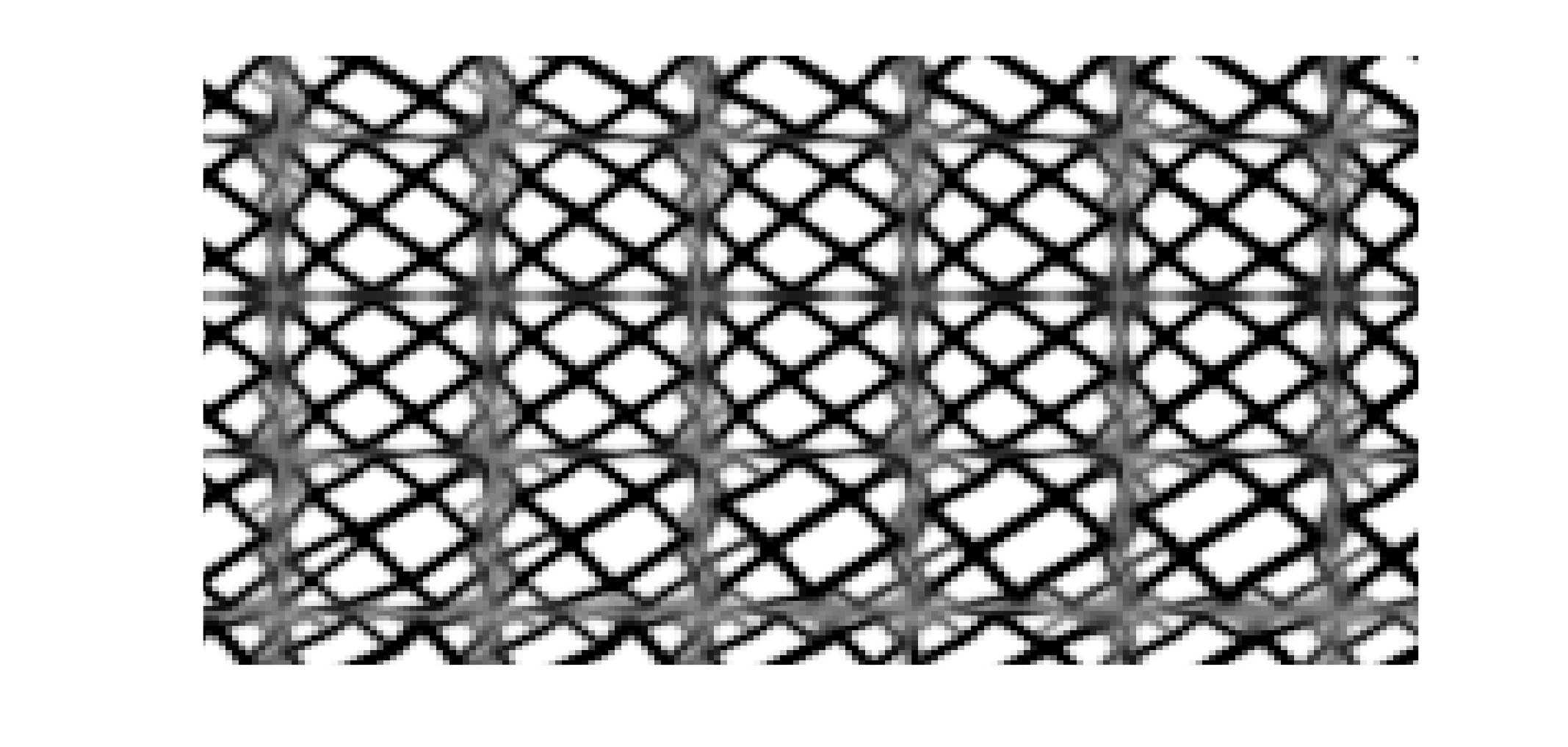}\label{fig9b}}\\
    \subfigure{\includegraphics[trim=0pt 40pt 0pt 400pt, clip=true, scale=0.35]{DensityGreyScale.eps} \notag}
\end{center}
\caption{Example 1, high-definition relative density distribution with \([\overline \rho_{min},\overline \rho_{max}]=[0.12,0.88]\).}\label{fig9}
\end{figure}

The result obtained after the fine-scale topology optimization process is applied to each of the cells is shown in Figure \ref{fig9a}. The stress paths defined by the cross-like distribution of densities \(\rho_e^{(k)}\) into each cell in the region around the neutral axis are similar to those shown in \cite[Fig. 16(b)]{Kumar2020} for the same problem and to the results in \cite[Fig. 3(b)]{Wu2018} for a cantilever beam loaded with a point load at the free end. Note that these paths change their slope becoming more horizontal as the distance to the neutral axis increases, i.e. as the ratio of the shear stress to the normal stress decreases.\par

Note that the applied load is skew-symmetric with respect the neutral axis, producing a material distribution mainly symmetrical. This is because the solution of the optimization problem seeks the minimum of the quadratic form $c(\rho)$, which is proportional to the strain energy density. Thus, there is no distinction between tension and compression in the final evaluation of stress.\par

As shown in Figure \ref{fig9b}, acceptably-continuous inter-cell material distribution is obtained at the fine scale. Despite this, since the lateral tractions act on the contour of each cell, the numerical behaviour of the fine-scale problem produces a frame effect over each cell, leading to a grid-like distribution. The elimination of these artefacts will be object of future research.\par

\begin{figure}[h! t]
\begin{center}
   \subfigure[{$[0.06,0.94]$}, Stage 6]{\includegraphics[trim=54pt 80pt 30pt 30pt, clip=true, scale=0.52]{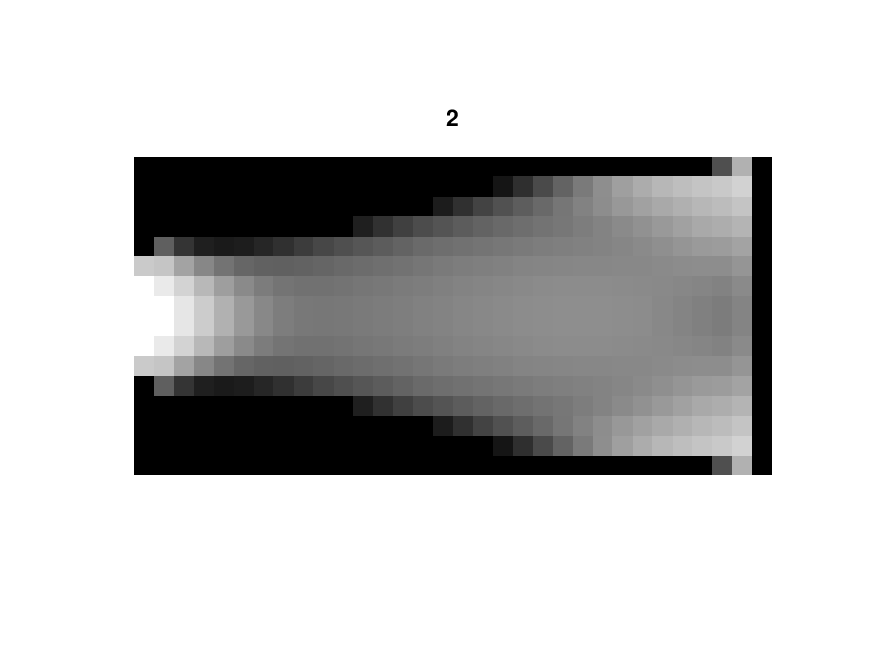}\label{fig10a}}\hskip 3pt
   \subfigure[{$[0.06,0.94]$}, Final solution]{\includegraphics[trim=60pt 30pt 40pt 30pt, clip=true, scale=0.2]{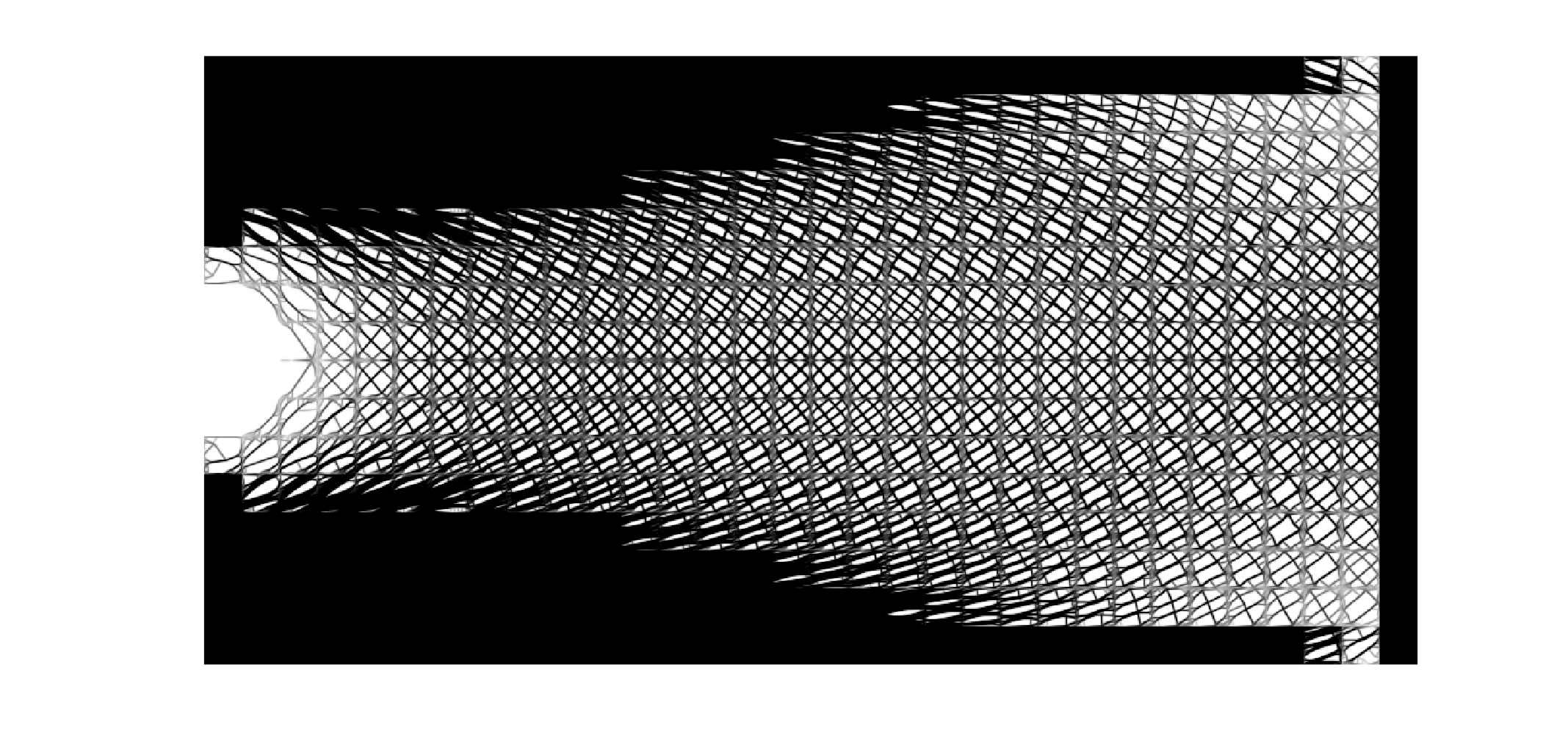}\label{fig10b}}\\
  \subfigure[{$[0.25,0.75]$}, Stage 7]{\includegraphics[trim=54pt 80pt 30pt 30pt, clip=true, scale=0.52]{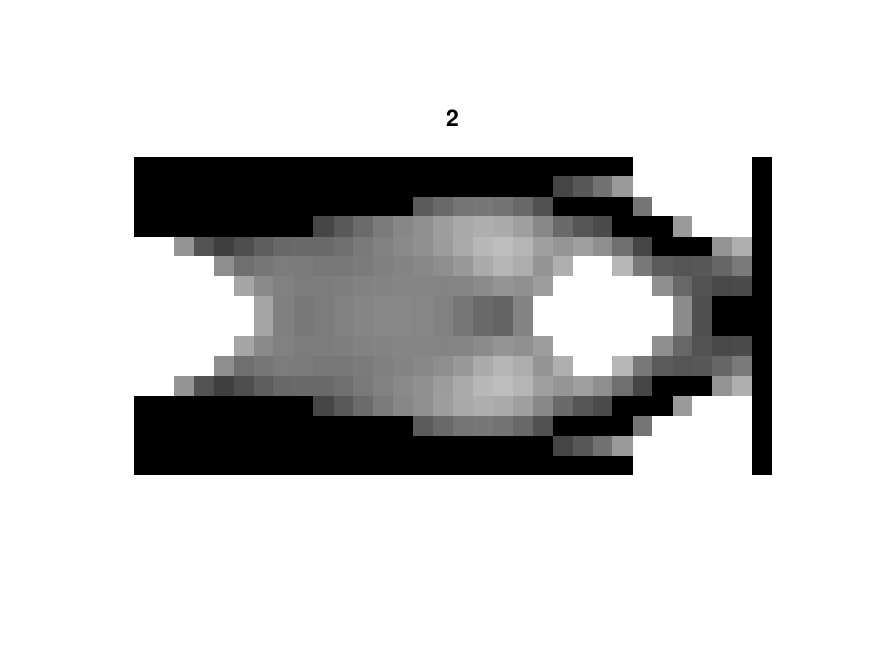}\label{fig10c}}\hskip 3pt
  \subfigure[{$[0.25,0.75]$}, Final solution]{\includegraphics[trim=60pt 30pt 40pt 30pt, clip=true, scale=0.2]{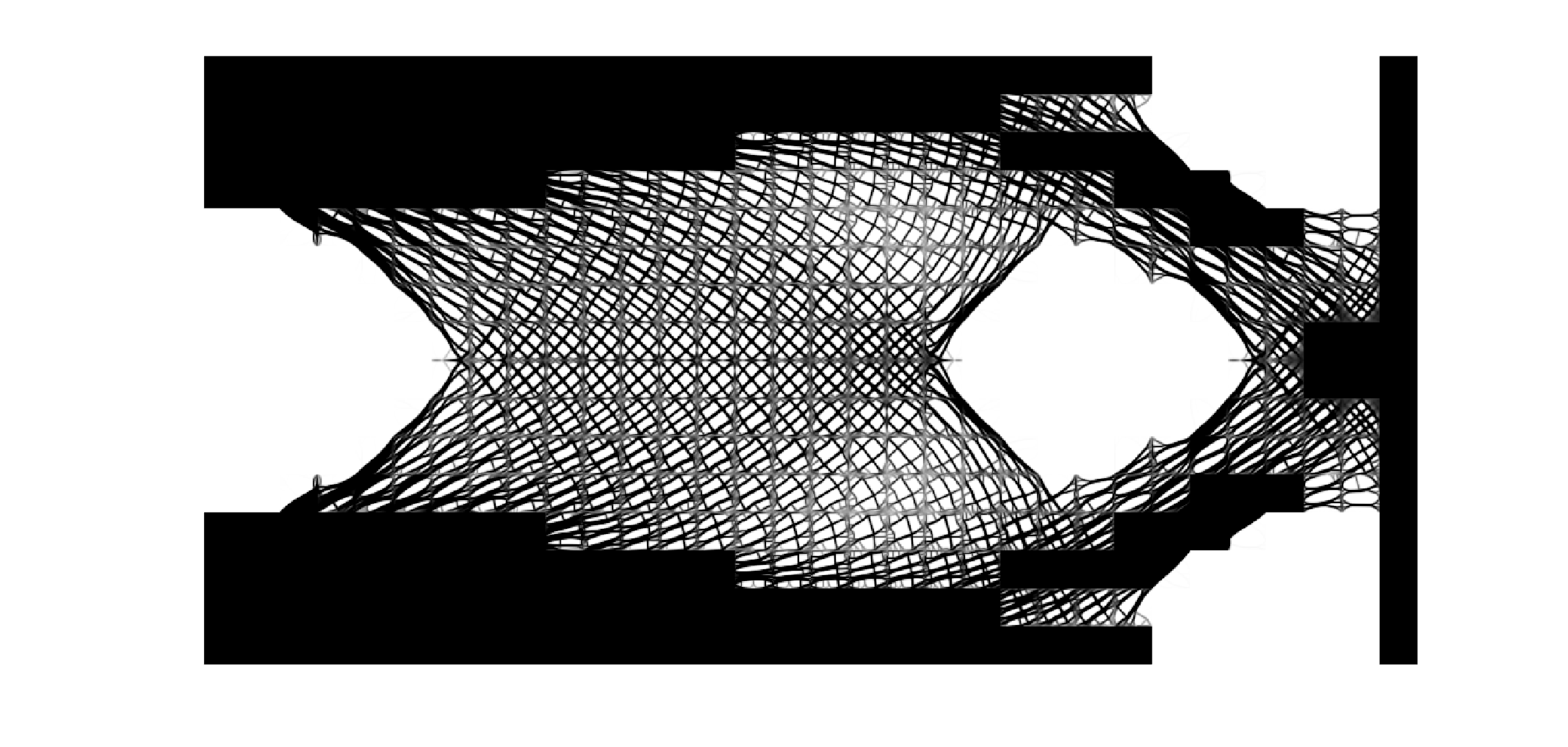}\label{fig10d}}\\
  \subfigure[{$[0.30,0.70]$}, Stage 13]{\includegraphics[trim=54pt 80pt 30pt 30pt, clip=true, scale=0.52]{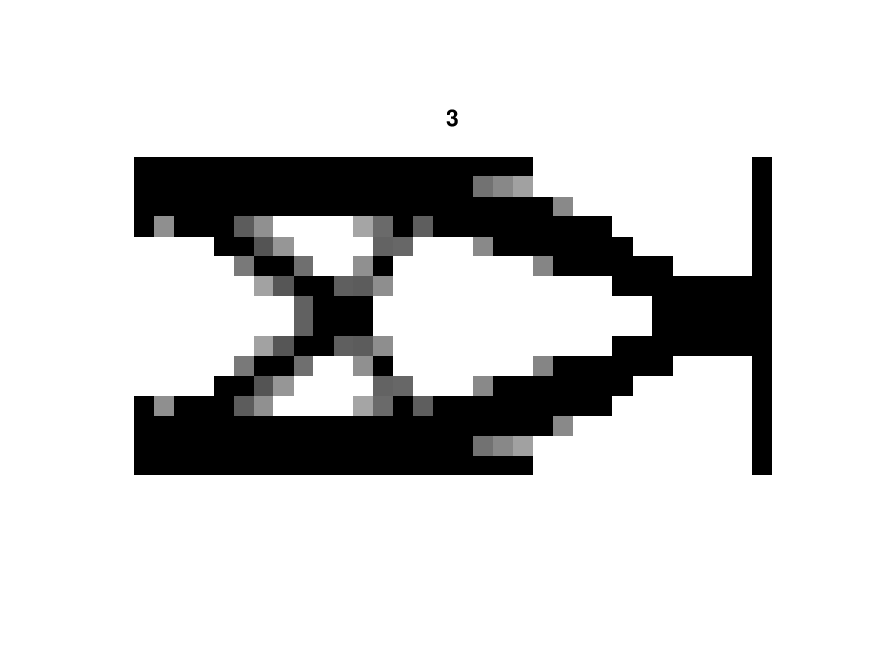}\label{fig10e}}\hskip 3pt
  \subfigure[{$[0.30,0.70]$}, Final solution]{\includegraphics[trim=60pt 30pt 40pt 30pt, clip=true, scale=0.2]{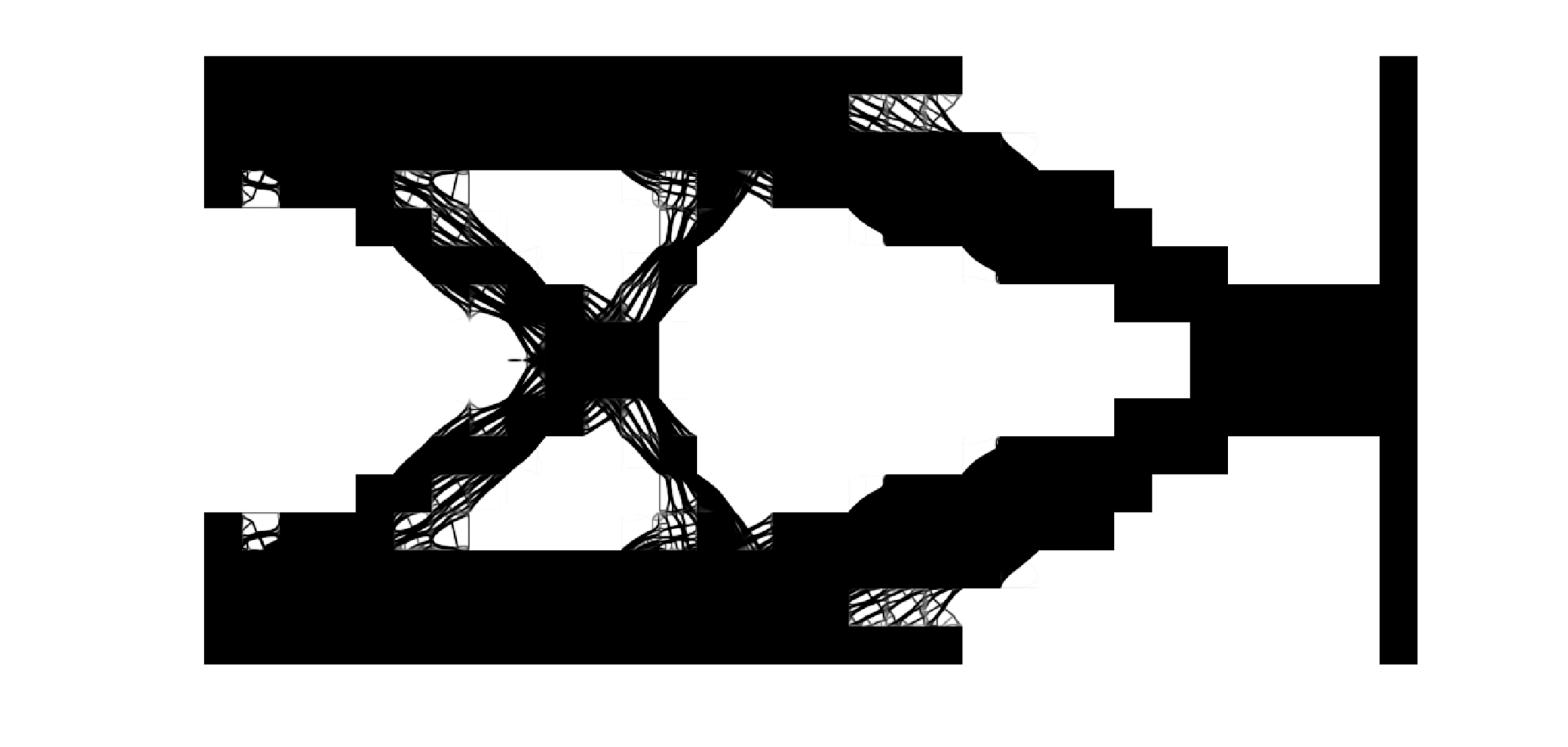}\label{fig10f}}\\
  \subfigure{\includegraphics[trim=0pt 40pt 0pt 400pt, clip=true, scale=0.35]{DensityGreyScale.eps} \notag}\\
\end{center}
\caption{Example 1, effect of different density thresholds on the relative density distribution.}\label{fig10}
\end{figure}

Next, the influence of density thresholds \([\overline \rho_{min},\overline \rho_{max}]\) on the final solution is studied. Three coarse-scale converged solutions with different relative density ranges have been represented in Figures \ref{fig10a}, \ref{fig10c}, \ref{fig10e} together with the corresponding fine-scale solutions in Figures \ref{fig10b}, \ref{fig10d}, \ref{fig10f}. As shown, the solution becomes closer to the classical optimization problem as the density range narrows (see \cite{Wu2018,Allaire2005,Dede2012} as examples). Accordingly, Figures \ref{fig10e} and \ref{fig10f} represent an almost solid-and-void solution for the narrowest interval, where only a few number of cells are optimized at the fine scale.\par

\begin{figure}[h! t]
\begin{center}
   \subfigure[{$[0.06,0.70]$}, Stage 5]{\includegraphics[trim=54pt 80pt 30pt 30pt, clip=true, scale=0.52]{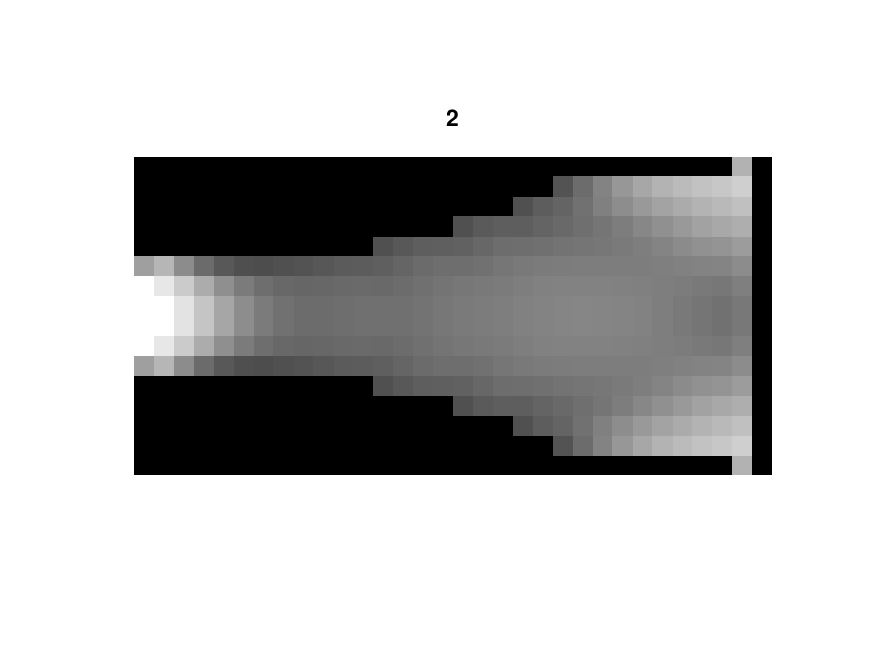}\label{fig11a}}\hskip 3pt
   \subfigure[{$[0.06,0.70]$}, Final solution]{\includegraphics[trim=60pt 30pt 40pt 30pt, clip=true, scale=0.2]{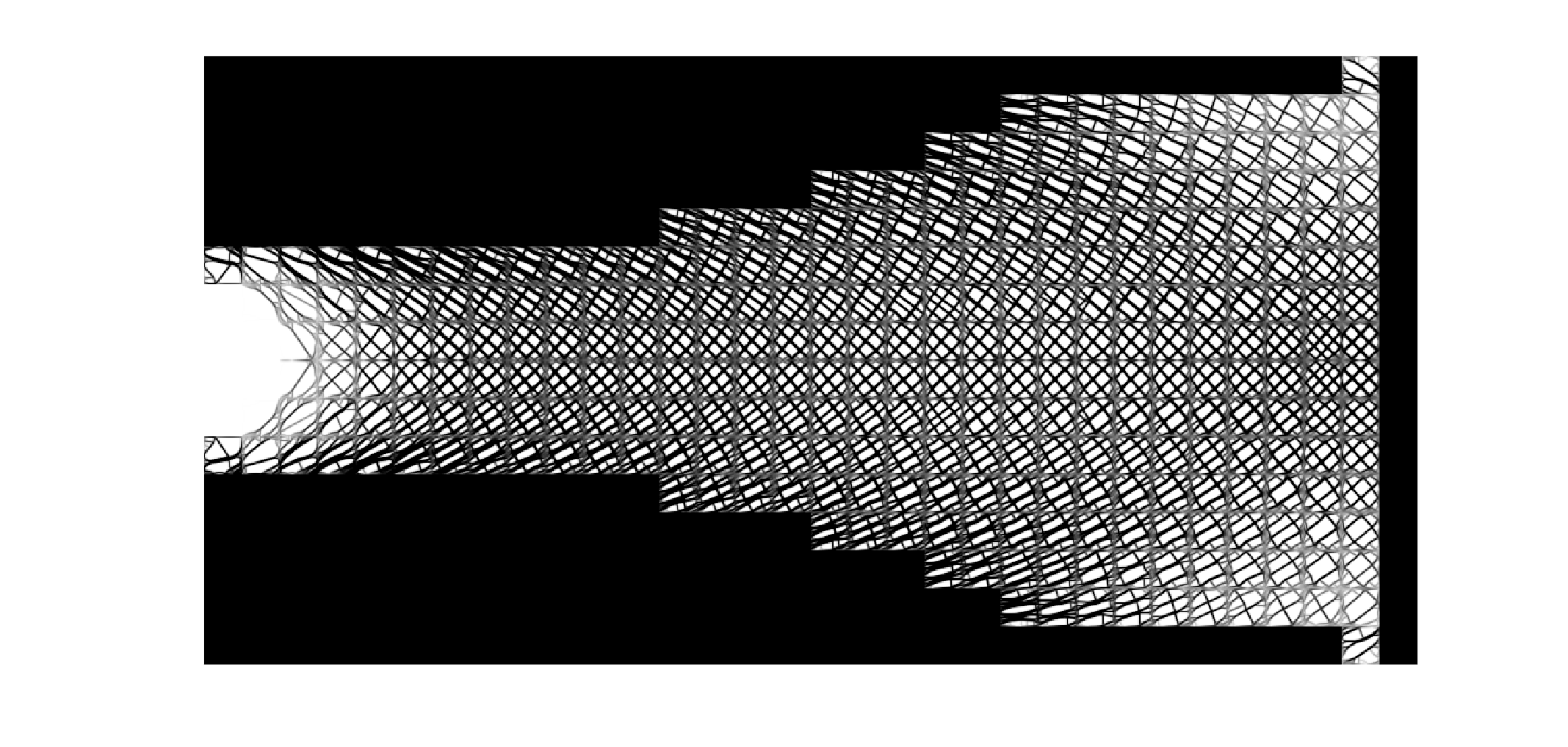}\label{fig11b}}\\
  \subfigure[{$[0.06,0.94]$}, Stage 6]{\includegraphics[trim=54pt 80pt 30pt 30pt, clip=true, scale=0.52]{Cantilever_006_094_Macro_IT6.eps}\label{fig11c}}\hskip 3pt
  \subfigure[{$[0.06,0.94]$}, Final solution]{\includegraphics[trim=60pt 30pt 40pt 30pt, clip=true, scale=0.2]{Cantilever_006_094_Micro.eps}\label{fig11d}}\\
  \subfigure[{$[0.30,0.94]$}, Stage 12]{\includegraphics[trim=54pt 80pt 30pt 30pt, clip=true, scale=0.52]{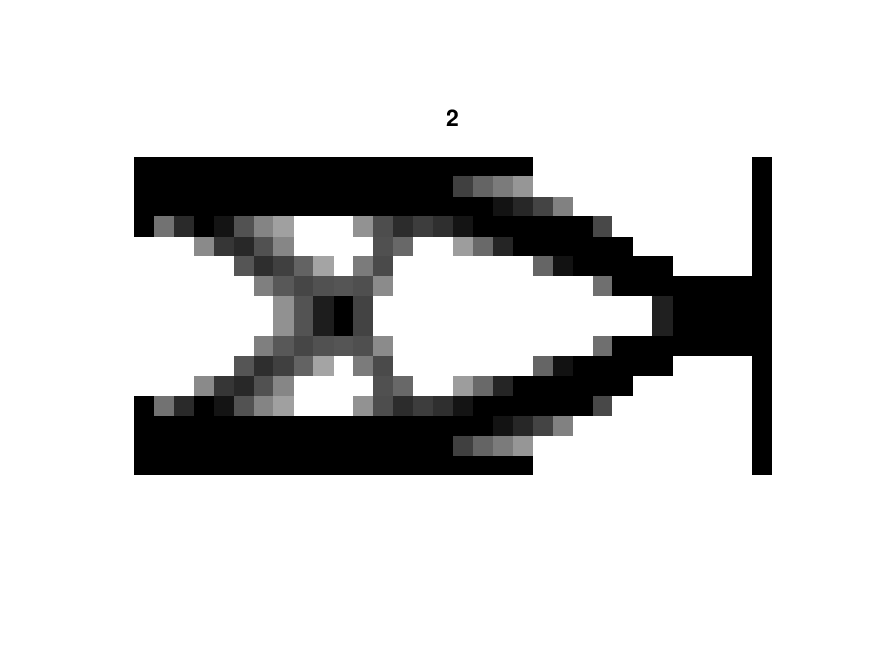}\label{fig11e}}\hskip 3pt
  \subfigure[{$[0.30,0.94]$}, Final solution]{\includegraphics[trim=60pt 30pt 40pt 30pt, clip=true, scale=0.2]{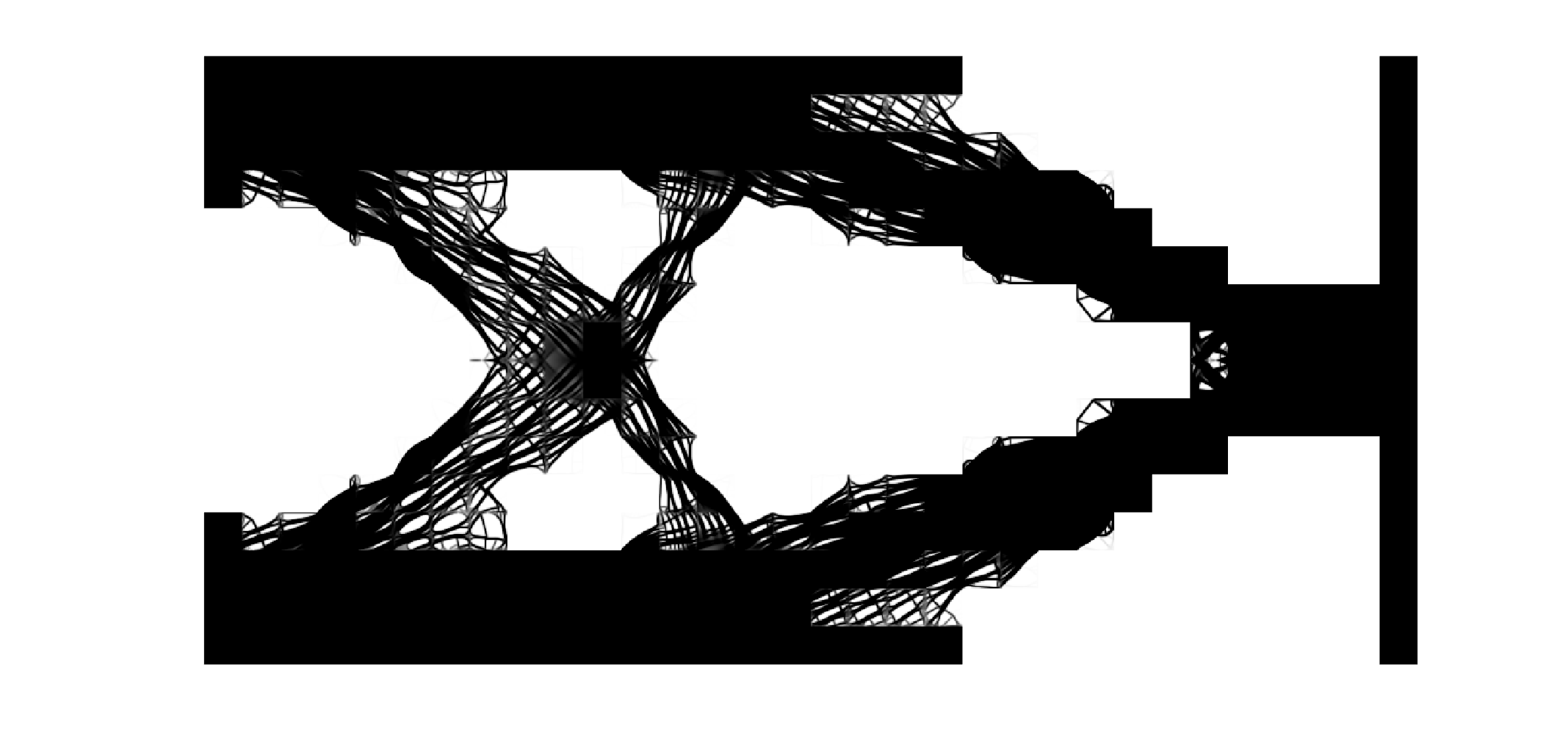}\label{fig11f}}\\
  \subfigure{\includegraphics[trim=0pt 40pt 0pt 400pt, clip=true, scale=0.35]{DensityGreyScale.eps} \notag}
\end{center}
\caption{Example 1, effect of non-symmetric density ranges on the relative density distribution.}\label{fig11}
\end{figure}

In addition, the effect of using non-symmetric density ranges is shown in Figure \ref{fig11} by comparing results from a symmetric gap (Figures \ref{fig11c} and \ref{fig11d}) with other from non-symmetric ones. Comparing Figures \ref{fig11a} and \ref{fig11c}, if only the upper density threshold $\overline \rho_{max}$ is reduced, the size of the solid region is increased, although not remarkably. However, comparing \ref{fig11c} and \ref{fig11e}, if only the lower density threshold $\overline \rho_{min}$ is increased to the same extend, the size of the void regions is increased significantly. Therefore, we can observe that the effect of reducing $\overline \rho_{max}$ is qualitatively smaller than that of increasing $\overline \rho_{min}$ because there are large areas with a relatively low value of density in this problem. Despite this, the effects of modifying either $\overline \rho_{min}$ or $\overline \rho_{max}$ are not completely uncoupled, since the total amount of material into the domain should be kept equal to $\overline \rho_0$ and the whole density distribution is affected by modifying one of the two ends of the range.

\subsection{Example 2. Cantilevered L-shape domain under bending load}
\begin{figure}[h! t]
\begin{center}
    \includegraphics[trim=0pt 20pt 0pt 0pt, clip=false,scale=0.5]{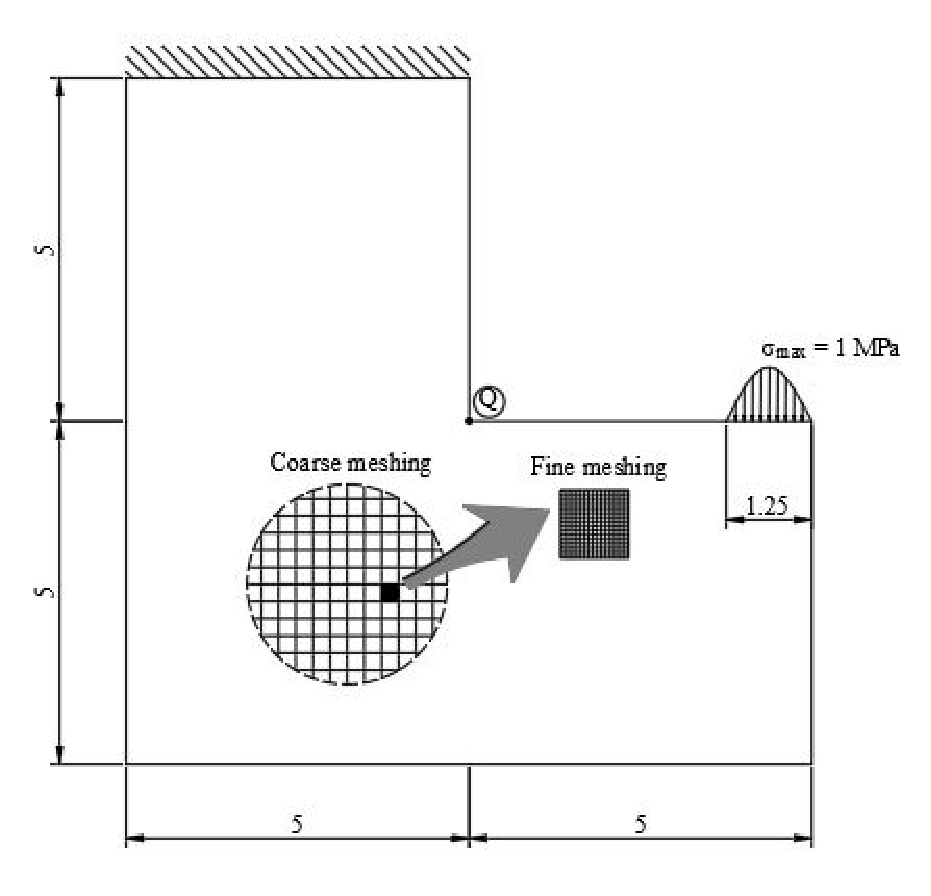}
\end{center}
\caption{Example 2, external loads and meshing.}\label{fig12}
\end{figure}

Our second example is a cantilevered L-shape domain under a parabolic vertical load, as shown in Figure \ref{fig12}, that produces a stress singularity at point Q. Outer dimensions of the domain are $10 \times 10$ m. The prescribed volume fraction is $\rho_0 = 0.5$. Other input parameters are outlined in Table \ref{tab2}.\par

\begin{table}[h! t]
\begin{center}
\begin{tabular}{lll}
\hline
Coarse scale & Fine scale & Projection\\
\hline
$32 \times 32$ elems. & $32 \times 32$ elems. & $\beta_0 = 1$\\
$p=1$  & $p=3$ & $\beta_{max} = 4.5$\\
$r_{min} = 1.5$  & $r_{min} = 1.3$ & $\mu = 0.5$\\
$\varepsilon = 0.03$ & $\varepsilon = 0.01$ & $M_{nd,min} = 45\%$\\
\hline
\end{tabular}
\end{center}
\caption{Input parameters for example 2.}\label{tab2}
\end{table}

\begin{figure}[h! t]
\begin{center}
  \subfigure[Stage 1]{\includegraphics[trim=80pt 30pt 80pt 0pt, clip=true, scale=0.55]{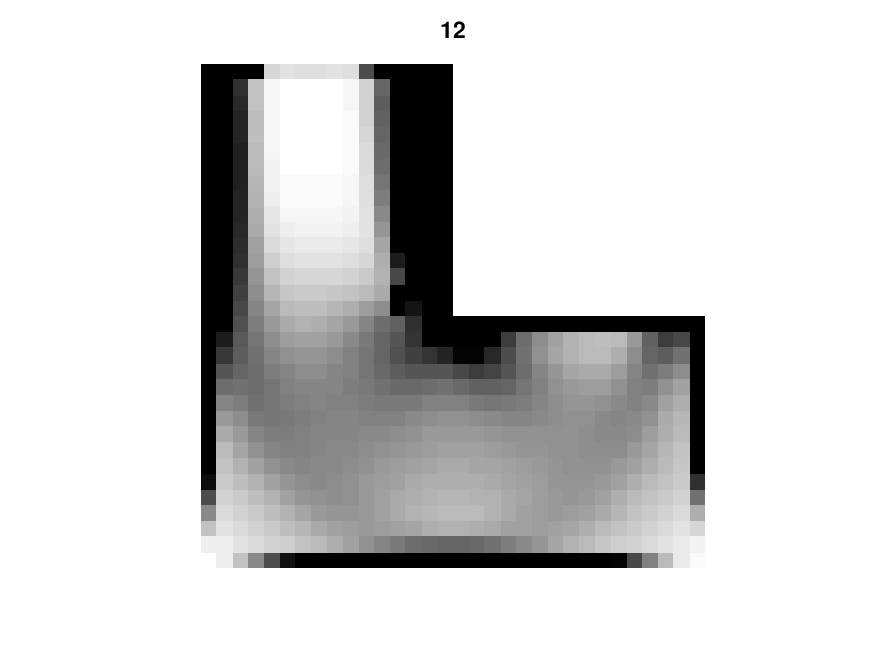}\label{fig13a}} \hskip 3pt
  \subfigure[Stage 3]{\includegraphics[trim=80pt 30pt 80pt 0pt, clip=true, scale=0.55]{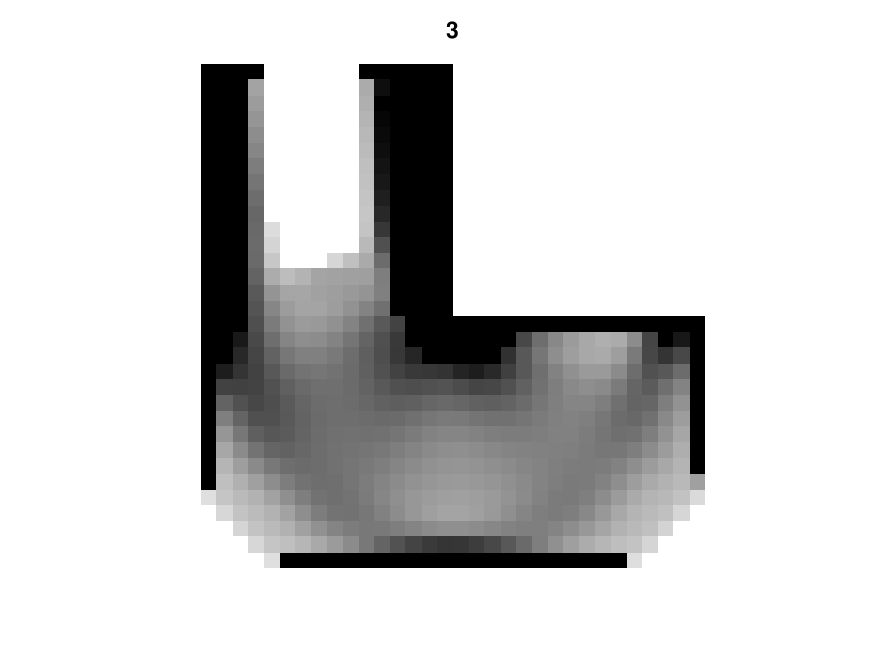}\label{fig13b}} \hskip 3pt
  \subfigure[Stage 5]{\includegraphics[trim=80pt 30pt 80pt 0pt, clip=true, scale=0.55]{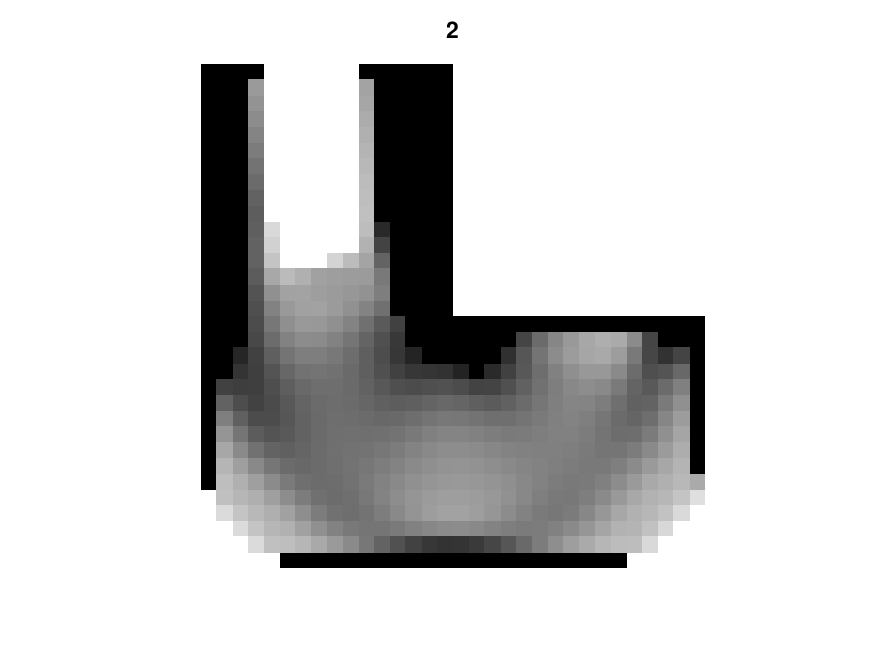}\label{fig13c}}\\
  \subfigure{\includegraphics[trim=0pt 40pt 0pt 400pt, clip=true, scale=0.35]{DensityGreyScale.eps} \notag}
\end{center}
\caption{Example 2, density distribution of the coarse problem with \([\overline \rho_{min},\overline \rho_{max}] [0.12,0.88]\).} \label{fig13}
\end{figure}

A first test with \([\overline \rho_{min},\overline \rho_{max}]=[0.12,0.88]\) is presented considered. Several iterations of the coarse-scale iterative procedure are plotted in Figure \ref{fig13}, where Fig. \ref{fig13c} represents the converged solution.\par

Due to the bending effect of the external load over the L-shaped domain, significant accumulations of material are noticed on the outer sides of the left upper region of the domain. In addition, remarkable accumulations can be observed around the singularity at point Q and near the boundary where the external load is applied. Even into the light-gray area, a slightly darker arch can be distinguished indicating the curved load path from the external load to the clamped end. Otherwise, void regions appear in the central zone of the clamped end and around the bottom corners of the domain, due to the reduced stress levels on these regions.\par

\begin{figure}[h! t]
\begin{center}
    \subfigure{\includegraphics[trim=0pt 50pt 0pt 0pt, clip=true, scale=0.55]{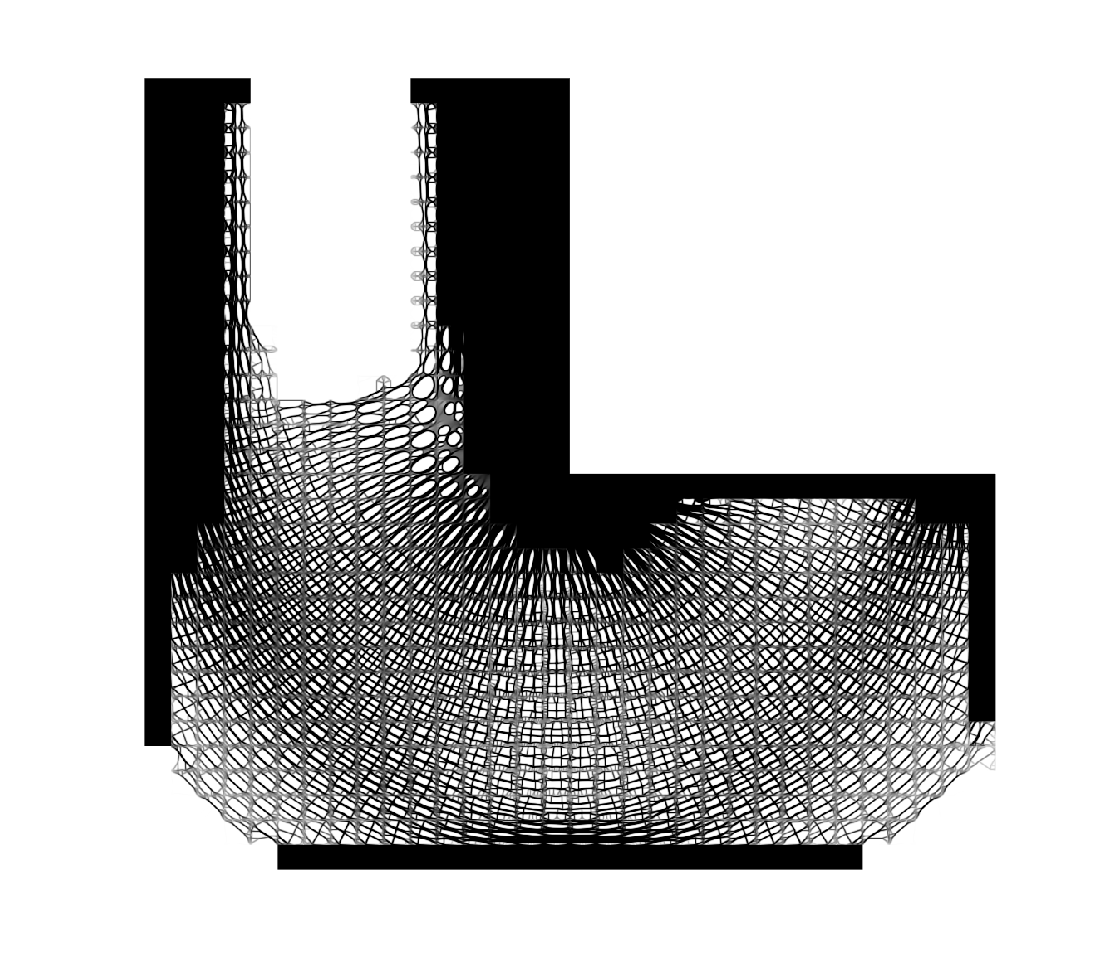}\notag}\\
    \subfigure{\includegraphics[trim=0pt 40pt 0pt 400pt, clip=true, scale=0.35]{DensityGreyScale.eps}\notag}
\end{center}
\caption{Example 2, high-definition relative density distribution with \([\overline \rho_{min},\overline \rho_{max}]=[0.12,0.88]\).}\label{fig14}
\end{figure}

After the coarse problem has been solved, the solution of the fine-scale problem is carried out (see Figure \ref{fig14}). As shown, paths of principal stresses in the lower region of the L-shape domain define arches and radii centered at point Q, as in \cite[Fig. 17]{Kumar2020}, \cite[Fig. 10]{Chandrasekhar2023}, \cite[Fig. 8]{Amstutz2010} and \cite[Tabs. 3,4]{Goncalves2022}. In the vertical branch of the beam these paths are nearly vertical because of the higher significance of normal stresses with respect to shear stresses.\par

\begin{figure}[h! t]
\begin{center}
   \subfigure[{$[0.06,0.94]$}, Stage 5]{\includegraphics[trim=80pt 30pt 80pt 0pt, clip=true, scale=0.55]{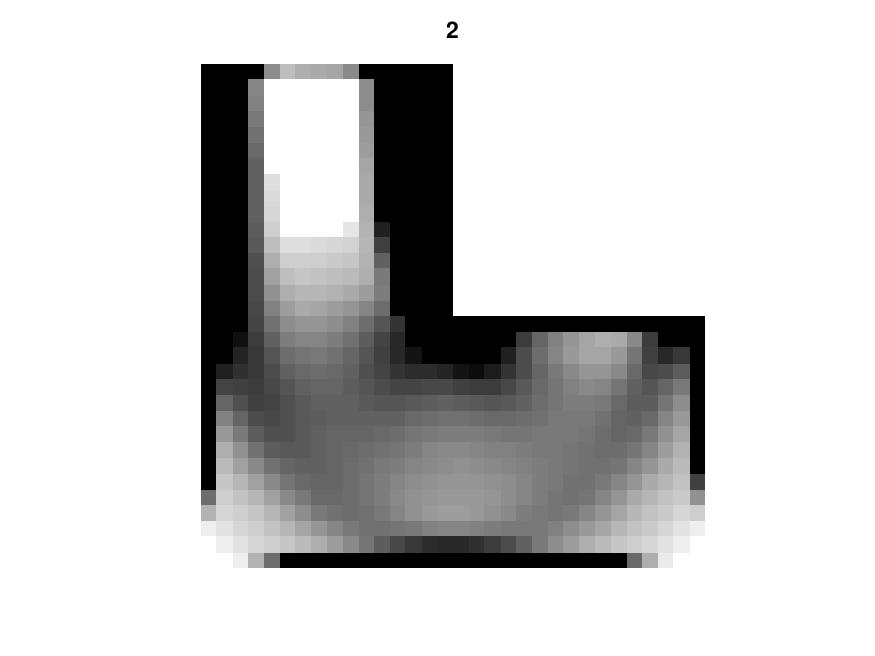}\label{fig15a}}\hskip 20pt
   \subfigure[{$[0.06,0.94]$}, Final solution]{\includegraphics[trim=40pt 30pt 40pt 0pt, clip=true, scale=0.375]{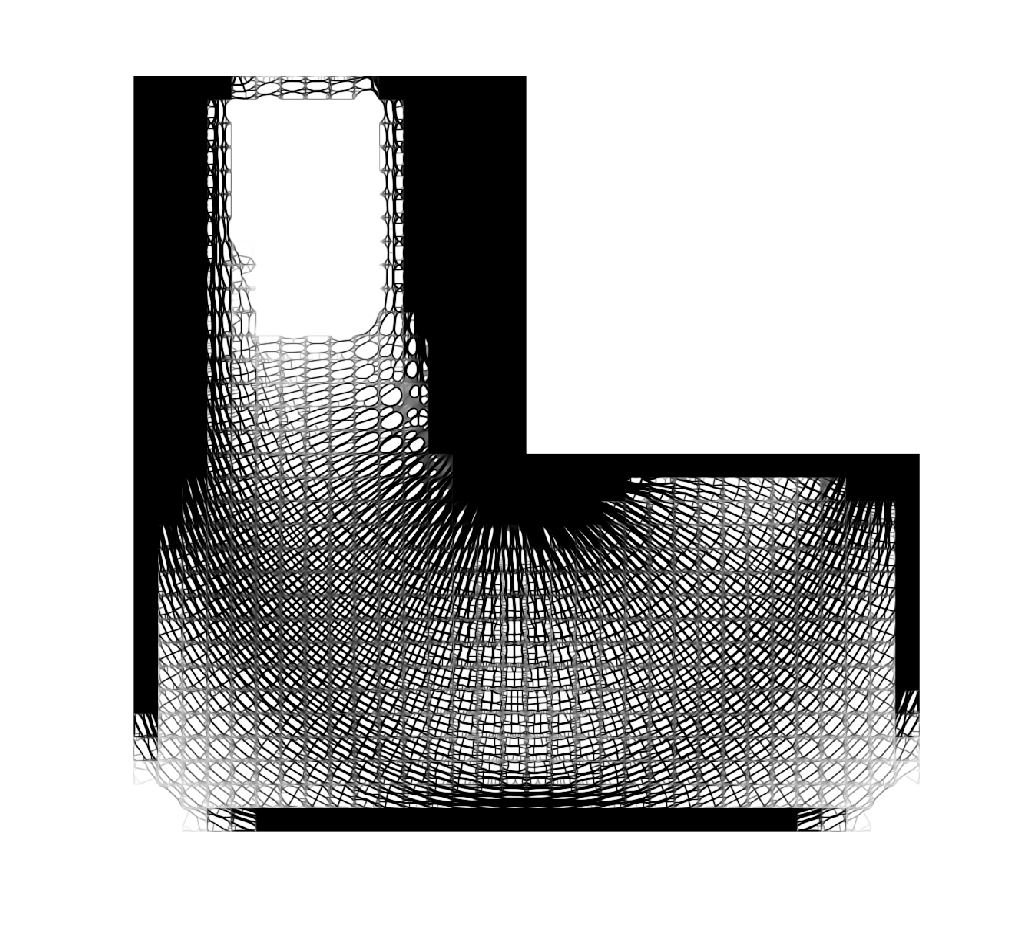}\label{fig15b}}\\
  \subfigure[{$[0.12,0.88]$}, Stage 5]{\includegraphics[trim=80pt 30pt 80pt 0pt, clip=true, scale=0.55]{Hook_012_088_Macro_IT5.eps}\label{fig15c}}\hskip 20pt
  \subfigure[{$[0.12,0.88]$}, Final solution]{\includegraphics[trim=40pt 30pt 40pt 0pt, clip=true, scale=0.35]{Hook_012_088_Micro.eps}\label{fig15d}}\\
  \subfigure[{$[0.30,0.70]$}, Stage 20]{\includegraphics[trim=80pt 30pt 80pt 0pt, clip=true, scale=0.57]{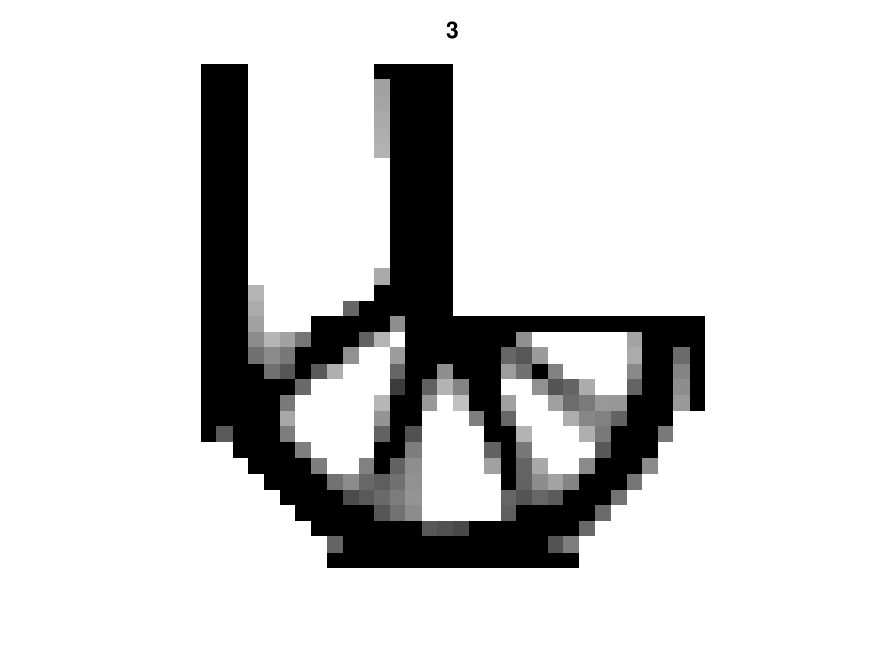}\label{fig15e}}\hskip 20pt
  \subfigure[{$[0.30,0.70]$}, Final solution]{\includegraphics[trim=40pt 30pt 40pt 0pt, clip=true, scale=0.365]{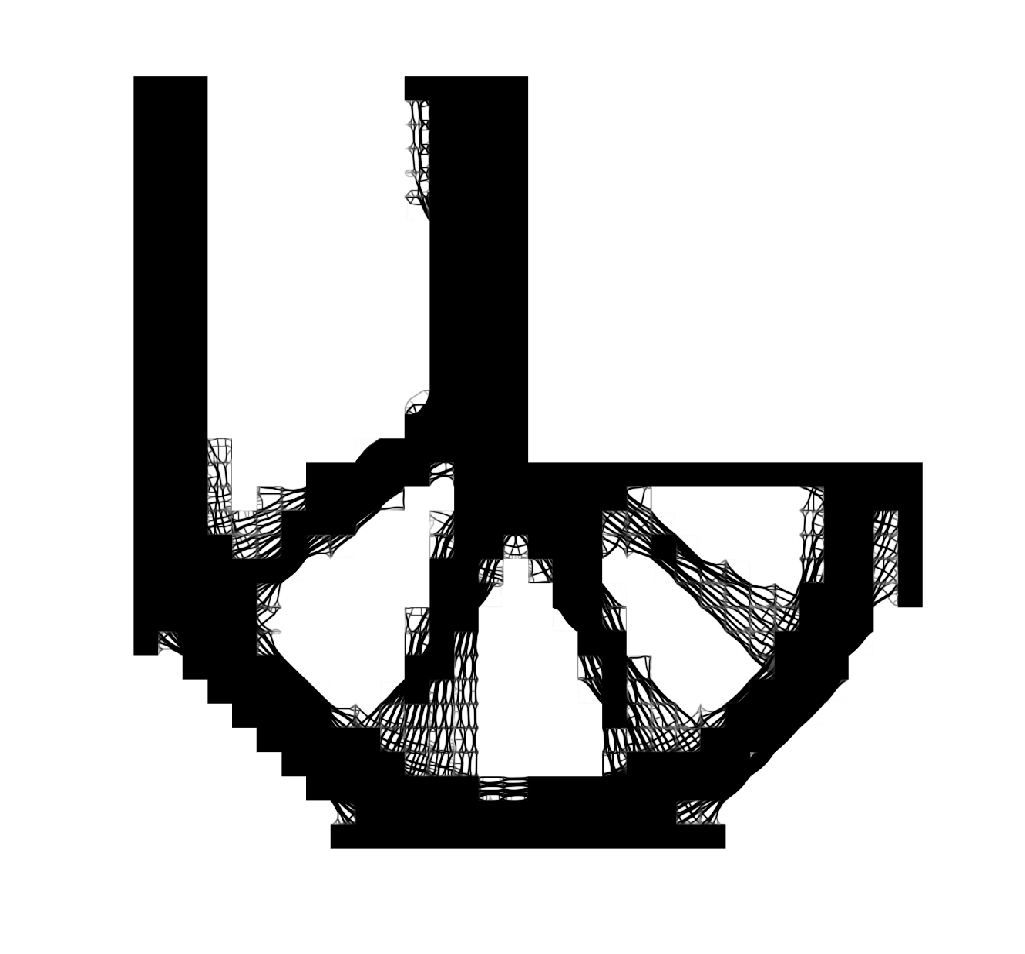}\label{fig15f}}\\
  \subfigure{\includegraphics[trim=0pt 40pt 0pt 400pt, clip=true, scale=0.35]{DensityGreyScale.eps}\notag}
\end{center}
\caption{Example 2, effect of different density thresholds on the relative density distribution.}\label{fig15}
\end{figure}

As for example 1, the influence of density thresholds on the final solution is studied considering three different ranges of density thresholds (see Figure \ref{fig15}). For each simulation, the converged coarse-scale (Figures \ref{fig15a}, \ref{fig15c}, \ref{fig15e}) and their respective fine-scale solutions (Figures \ref{fig15b}, \ref{fig15d}, \ref{fig15f}) are displayed. As in example 1, the smaller the range of density thresholds, the larger the total size of solid and void regions. Therefore, for very small intervals, the fine-scale solution will tend to the classical topology optimization solution (see \cite{Amstutz2010,Goncalves2022} as examples).\par

\begin{figure}[h! t]
\begin{center}
   \subfigure[{$[0.06,0.70]$}, Stage 10]{\includegraphics[trim=80pt 30pt 80pt 0pt, clip=true, scale=0.55]{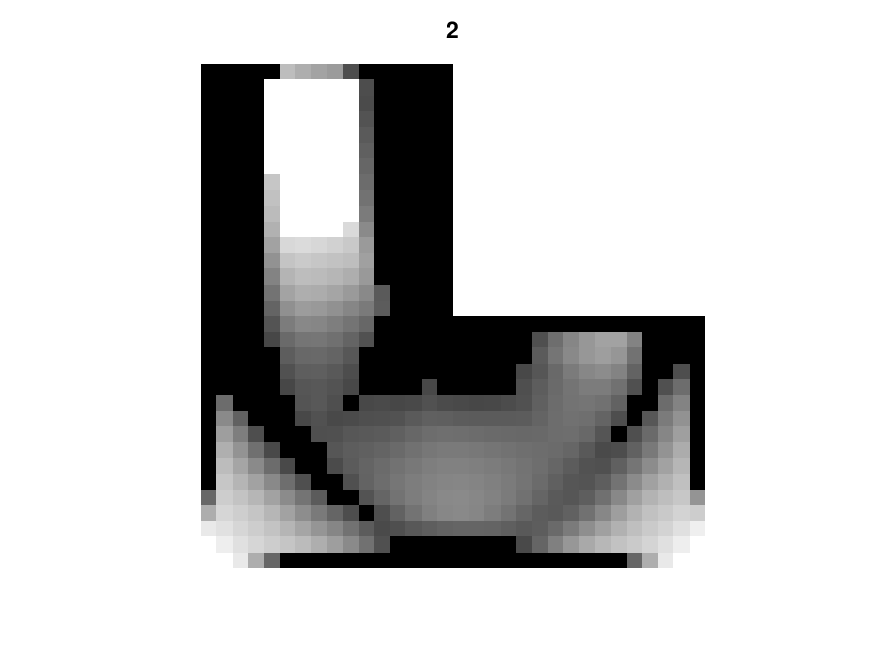}\label{fig16a}}\hskip 20pt
   \subfigure[{$[0.06,0.70]$}, Final solution]{\includegraphics[trim=40pt 30pt 40pt 0pt, clip=true, scale=0.36]{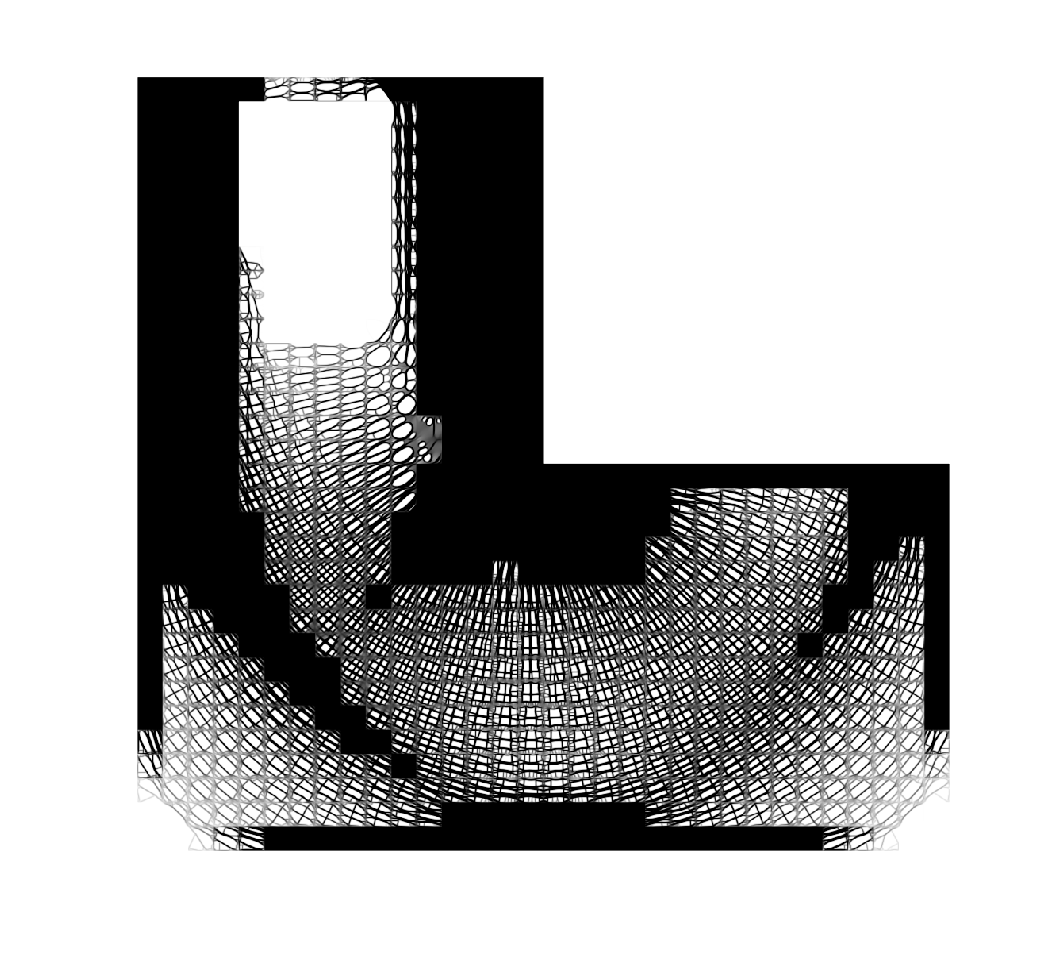}\label{fig16d}}\\
  \subfigure[{$[0.06,0.94]$}, Stage 5]{\includegraphics[trim=80pt 30pt 80pt 0pt, clip=true, scale=0.55]{Hook_006_094_Macro_IT5.eps}\label{fig16b}}\hskip 20pt
  \subfigure[{$[0.06,0.94]$}, Final solution]{\includegraphics[trim=40pt 30pt 40pt 0pt, clip=true, scale=0.365]{Hook_006_094_Micro.eps}\label{fig16e}}\\
  \subfigure[{$[0.30,0.94]$}, Stage 9]{\includegraphics[trim=80pt 30pt 80pt 0pt, clip=true, width = 150pt, height= 170pt]{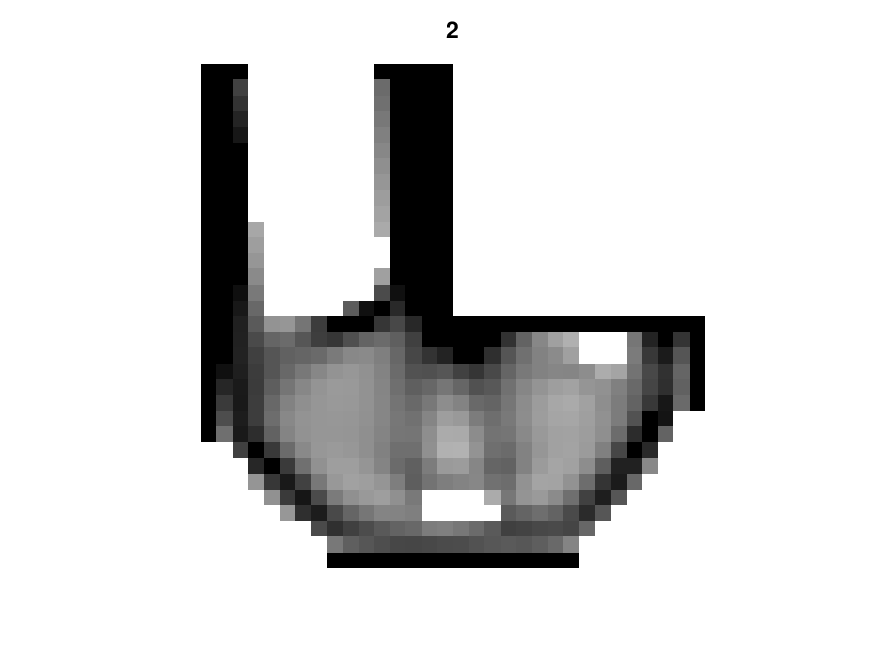}\label{fig16c}} \hskip 20pt
  \subfigure[{$[0.30,0.94]$}, Final solution]{\includegraphics[trim=40pt 30pt 40pt 0pt, clip=true, width = 150pt, height = 163pt]{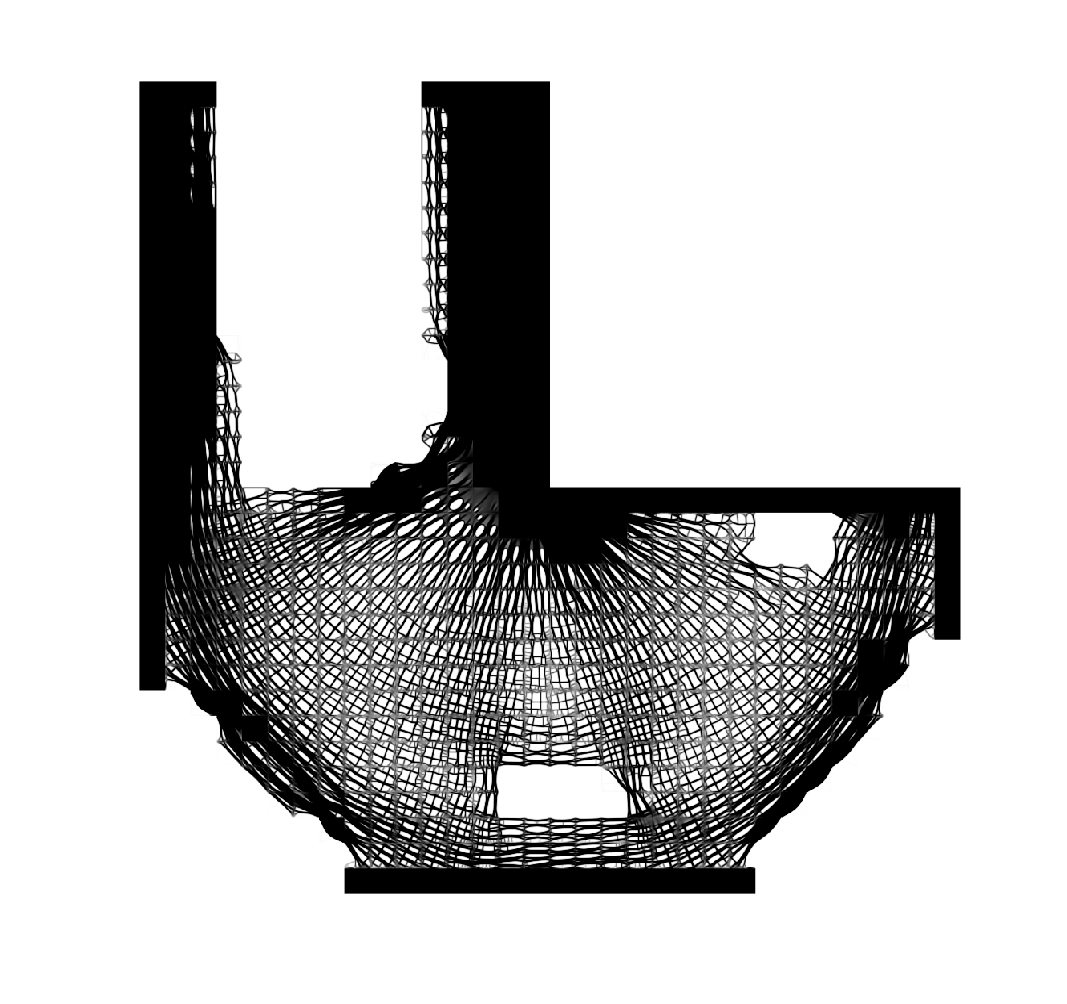}\label{fig16f}}\\
  \subfigure{\includegraphics[trim=0pt 40pt 0pt 400pt, clip=true, scale=0.35]{DensityGreyScale.eps}\notag}
\end{center}
\caption{Example 2, effect of non-symmetric density ranges on the relative density distribution.}\label{fig16}
\end{figure}

Additionally, simulations with non-symmetric gaps (Figures \ref{fig16a}, \ref{fig16b} , \ref{fig16e}, \ref{fig16f}) have been compared to results with a symmetric gap in Figures \ref{fig16b}, \ref{fig16c}, obtaining similar trends to those obtained for example 1, i.e., when $\overline \rho_{max}$ is reduced, the size of solid regions is incremented and when $\overline \rho_{min}$ is increased, void regions become larger too.\par

\section{Concluding remarks}\label{remarks}
In the present work, a two-level topology optimization technique based on the SIMP method is proposed. Firstly, a topology optimization problem is solved at the coarse level by using the FEM, which determines the general distribution of material in the component. As a result, a constant relative density value is assigned to each coarse-scale cell represented by one finite element. A range of density thresholds have been introduced to avoid further optimization of cells with extreme relative densities. Secondly, after evaluating a continuous traction field by means of the traction equilibrating method, a new topology optimization problem is solved into each cell at the fine level. Finally, after compiling the solution of each individual cell into one single result, a reasonably continuous across-cells material distribution is obtained. The ability of the proposed methodology to uncouple the topology optimization of each individual cell allows to significantly reduce the computational resources and computing time required to solve this kind of problems. In order to show the robustness of the method, two numerical examples have been presented: a cantilever beam under shear loading and a L-shape domain under bending load. Both have shown good agreement with other published works \cite{Kumar2020,Wu2018,Allaire2005,Dede2012,Chandrasekhar2023,Amstutz2010,Goncalves2022} and significant sharpness level.\par

The proposed methodology provides high-resolution solutions of the optimal distribution of material into the design domain with geometrical continuity through a two-level topology optimization method, where the continuity of tractions across the coarse-scale elements has been imposed. The novel application of the traction equilibrating method, based on the method described in \cite{Ladeveze1996}, to uncouple the topology optimization of each individual cell represents the key aspect of the proposed methodology.\par

This paper also presents a very simple methodology, based on the use of prescribed limiting values of relative density, to transform almost void (\(\rho_i^{(k)}<\overline \rho_{min}\)) or almost solid (\(\rho_i^{(k)}>\overline \rho_{max}\)) cells into completely void or completely solid cells in the final solution. This new methodology not only reduces the total number of cells on which to perform fine-scale optimization, but also reduces the manufacturing cost by avoiding the manufacturing of nearly void or nearly solid cells. In fact, making certain elements completely void or solid does not constrain sustantively the whole design of the component. The numerical results show that the narrower the range of density thresholds, the closer is the final result to the classic SIMP solutions, with a greater relative importance of the coarse scale compared to the fine scale in the final solution. In fact, narrow density ranges can produce noticeable deviations from the two-level optimum solution obtained with wide ranges. On the other hand, regarding the non-symmetry of the range of density thresholds, the presented solutions are qualitatively more sensitive to an increase of  $\overline \rho_{min}$ (larger void regions) than to a decrease of $\overline \rho_{max}$ (larger solid regions).\par

The main results presented in this paper show that the proposed methodology, once developed for 3D applications, has the potential to become a design tool for industrial environments, especially for additive manufacturing applications.

\section{Acknowledgements}\label{Acknowledgements}
The authors gratefully acknowledge the support provided by Generalitat Valenciana (Prometeo/2021/046), Vicerrectorado de Investigación de la Universitat Politècnica de València (PAID-11-22), Ministerio de Ciencia e Innovación, Agencia Estatal de Investigación and FEDER (PID2022-141512NB-I00), as well as by Ministerio de Educación y Formación Profesional (FPU19/02103 and FPU20/04828). Furthermore, the authors wish to thank Mikel Barral for his seminal contribution to this work.

\begin{appendix}
\section{Special cases of equilibrium}\label{Spcases}

\begin{figure}[h! t]
\begin{center}
  \subfigure[Sample of different cases]{\includegraphics[scale=0.4]{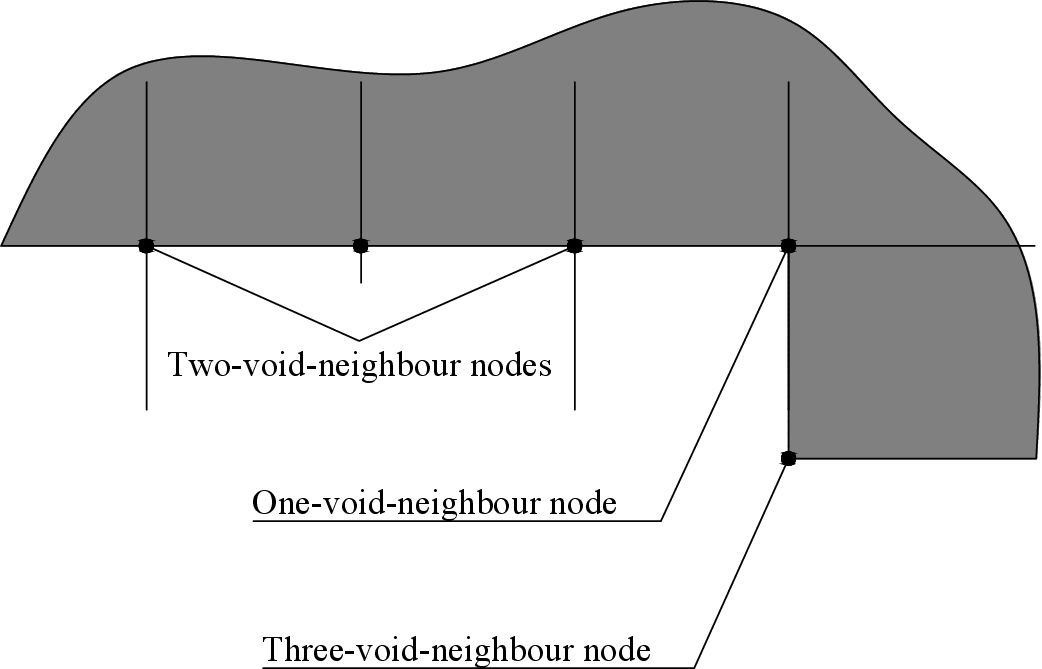}\label{figA1a}}\qquad
  \subfigure[Ordering of elements around the node]{\includegraphics[scale=0.28]{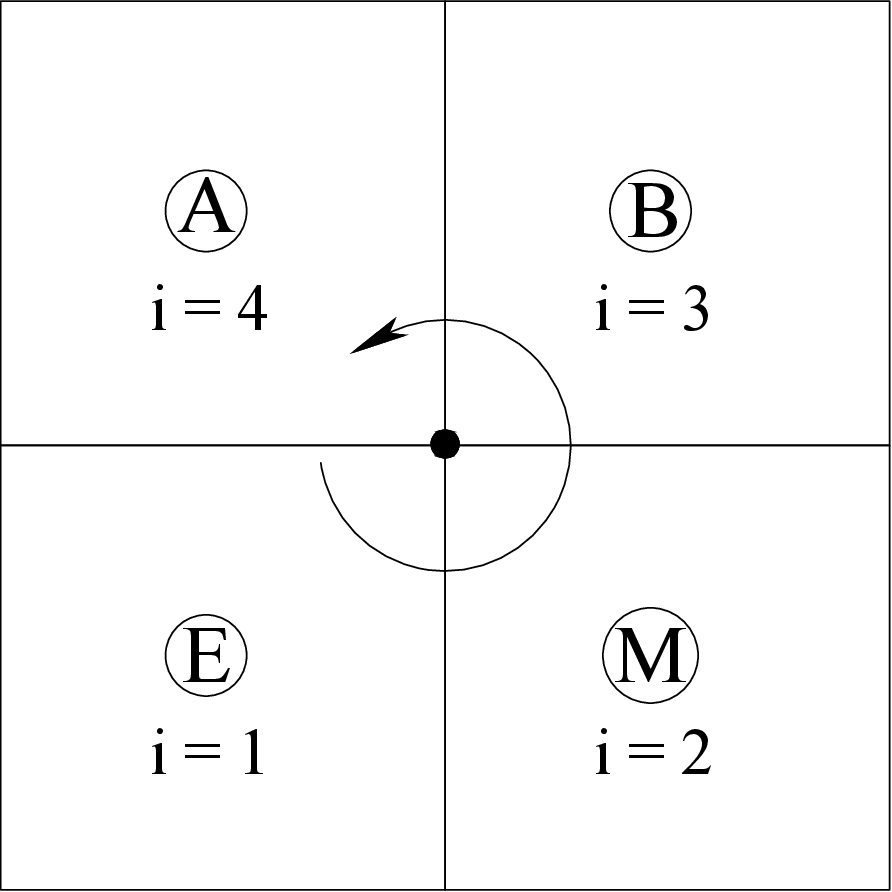}\label{figA1b}}\\
\end{center}
\caption{Special cases of nodes neighboring void elements.}\label{figA1}
\end{figure}

The internal nodes neighbouring elements which have been turned into void (assigning them $\rho = 10^{-3}$) in the coarse problem iterative procedure included in the algorithm described in Figure \ref{fig2}, are considered special cases of equilibrium. In fact, if one or more elements around a node become void, the edge between void and non-void elements becomes equivalent to a Neumann boundary subjected to null tractions (Figure \ref{figA1}). However, if the general forces polygon shown in Figure \ref{fig51a} is used and the \emph{pole} is kept at the geometrical mass centre of the polygon, it is possible than a vanishing nodal force could produce corresponding non-vanishing side forces, yielding undesired accumulations of material along the new Neumann-like boundary in the fine-scale optimization. In order to overcome this issue, further analysis is appropriate.\par

\begin{figure}[h! t]
\begin{center}
  \subfigure[One-void-neighbour element]{\includegraphics[trim=20pt 0pt 20pt 50pt, clip=false, scale=0.3]{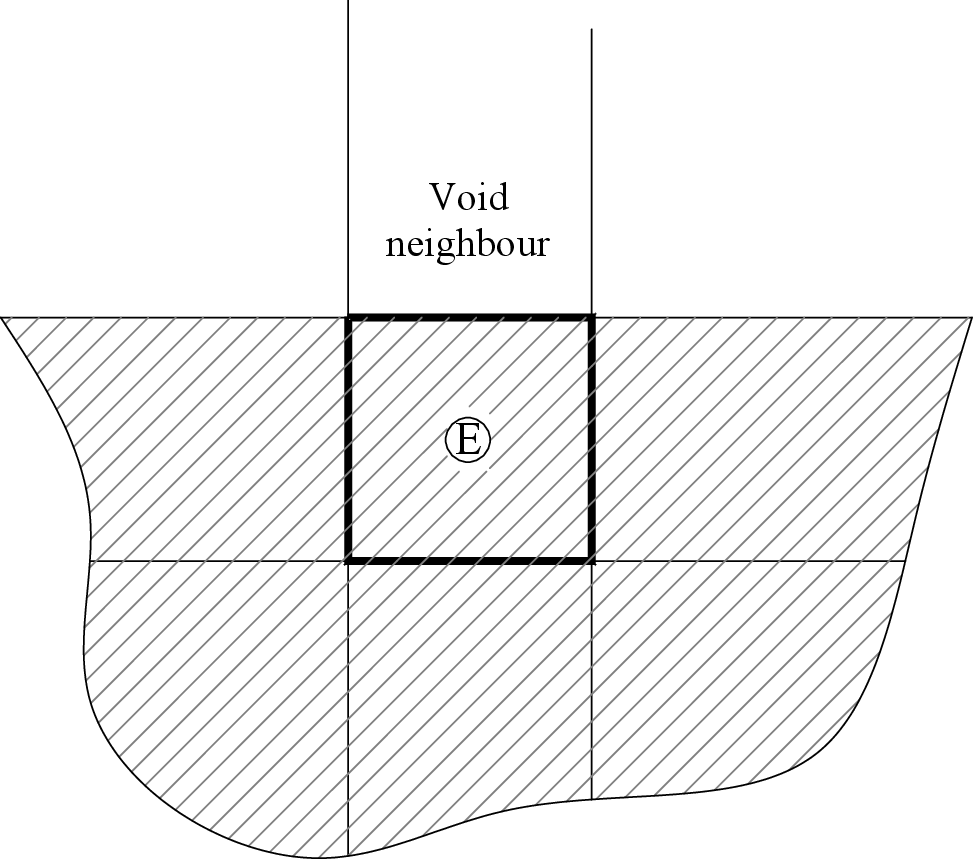}\label{figA2a}}\qquad
  \subfigure[Two-void-neighbour element]{\includegraphics[trim=20pt 10pt 20pt 50pt, clip=false, scale=0.29]{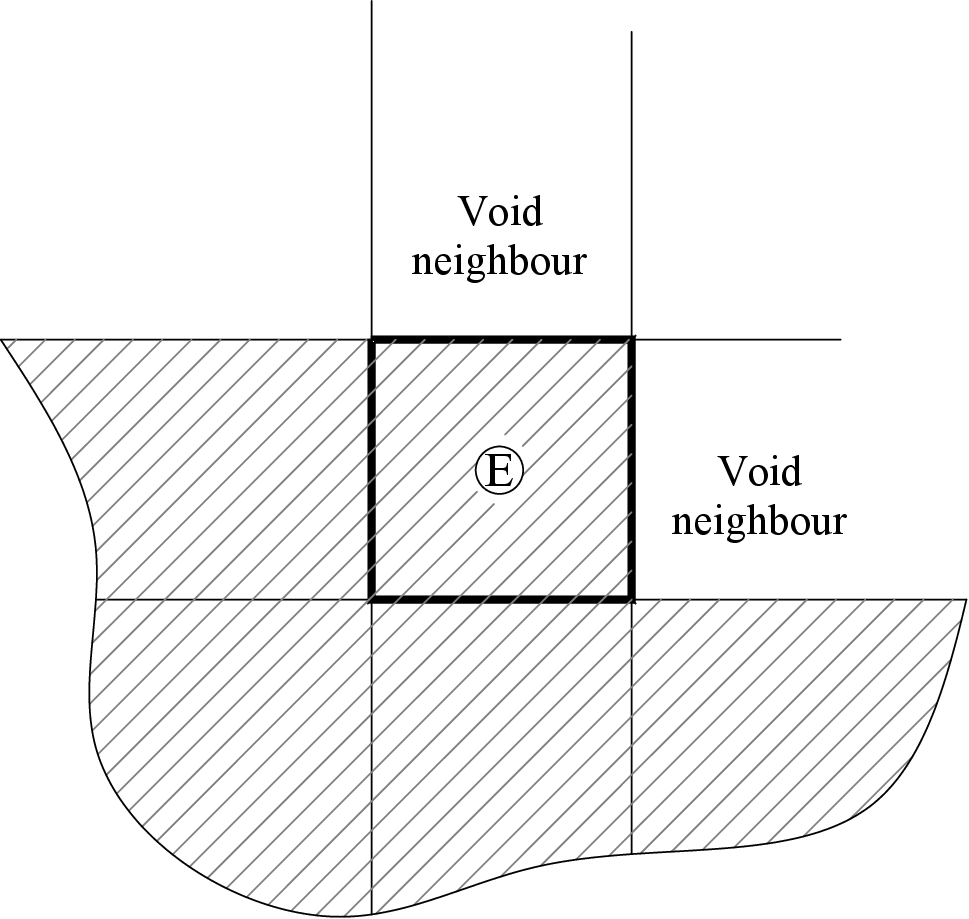}\label{figA2b}}\qquad
  \subfigure[Three-void-neighbour element]{\includegraphics[trim=30pt 20pt 20pt 50pt, clip=false, scale=0.275]{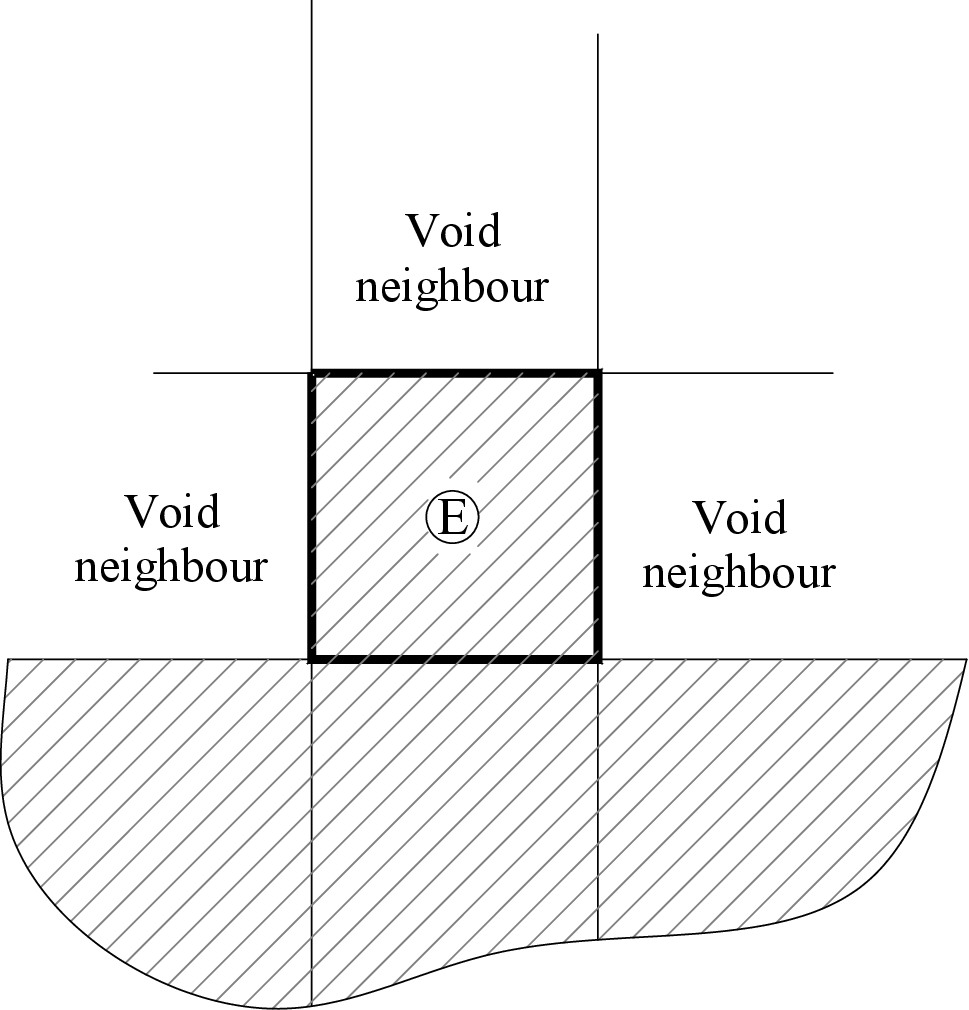}\label{figA2c}}
\end{center}
\caption{Cases of neighbouring through edges.}\label{figA2}
\end{figure}

Let us consider element $E$ to be optimized, included in the shaded area of Figure \ref{figA2}. It should be noted that only one or two void edge neighbours can be considered. Otherwise, if three edges of an ordinary element are neighbouring void elements, the global equilibrium of moments would not not feasible. In that case, element $E$ would be turned into void (even if its density is into the density range) in order to keep the whole equilibrium of the component.\par

Some examples of special cases involving void elements and the order of elements adopted in our implemented code to solve the equilibrium of nodes are represented in Figure \ref{figA1}. The particular features of each case are explained below.

\begin{figure}[h! t]
\begin{center}
  \subfigure[One-void-neighbour node]{\includegraphics[scale=0.32]{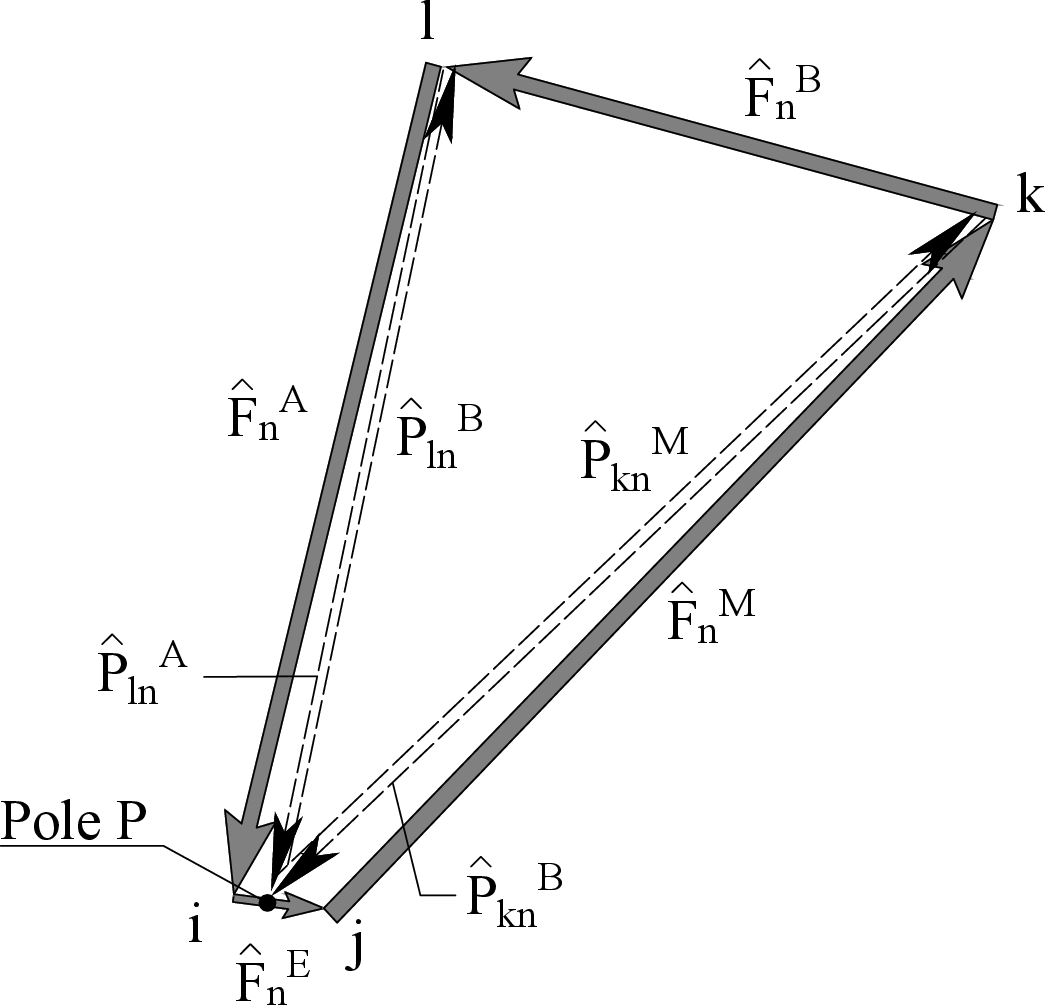}\label{figA3a}}\qquad
  \subfigure[Two-void-neighbour node]{\includegraphics[trim=-80pt 0pt -80pt 0pt, clip=true, scale=0.25]{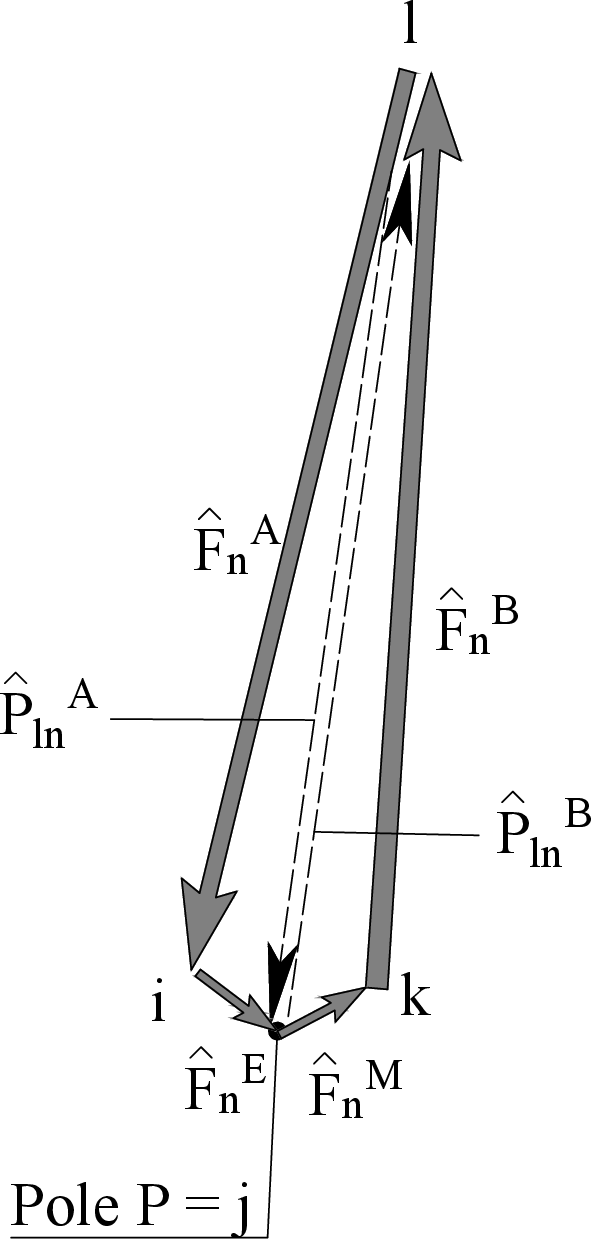}\label{figA3b}}\qquad
  \subfigure[Three-void-neighbour node]{\includegraphics[scale=0.3]{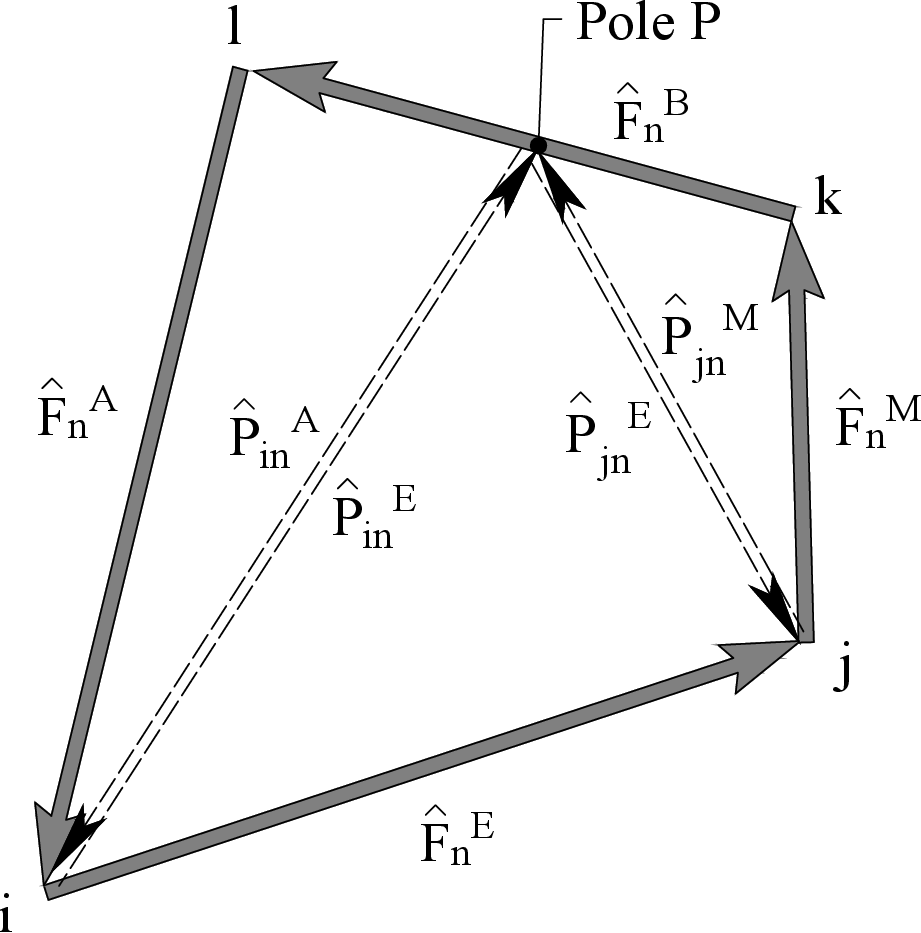}\label{figA3c}}\\
\end{center}
\caption{Special cases for Maxwell polygon.}\label{figA3}
\end{figure}

\begin{description}
     \item[One-void-neighbour node.] In this case, only a nodal force around the node will be significantly smaller than the rest of them. Therefore, the quadrilateral becomes similar to a triangle (see Figure \ref{figA3} as an example). The pole will be located at the midpoint of the vanishing nodal force in order to obtain negligible side forces associated to the node of the void element.
     \item[Two-void-neighbour node.] Nodes with two adjacent void elements around the node are included into this group. Thus, two consecutive (excluding the singular checkerboard patterns) nodal forces of the polygon are smaller than the other by several orders of magnitude and the quadrilateral becomes a needle-like polygon (see Figure \ref{figA3}). Extending the idea exposed in the previous case, the \emph{pole} is located at the vertex between the two neglecting nodal forces.
     \item[Three-void-neighbour node]. In the latter special case, nodes with three void elements around the node are considered. The polygon in this case is a quadrilateral of arbitrary shape. Here, all the forces are of small magnitude and similar to each other. In order to reduce the side forces of the non-void element ($B$ in the case represented in Figure \ref{figA3}), the \emph{pole} is located at the midpoint of its corresponding nodal force.
\end{description}

\end{appendix}


\bibliographystyle{elsarticle-num}
\bibliography{2LTopOpt}

\end{document}